\documentstyle[aps,epsfig,subfigure,cite]{revtex}
\begin{document}
\title{A scalar model of inhomogeneous elastic and granular media}
\author{M. L. Nguyen and S. N. Coppersmith}
\address{The James Franck Institute and Department of Physics\\
         The University of Chicago, 5640 S. Ellis Ave., Chicago, IL 60637}
\date{May 1, 2000}
\maketitle
\begin{abstract}
We investigate theoretically how the stress propagation characteristics
of granular materials evolve as they are subjected to increasing
pressures, comparing the results of a two-dimensional scalar lattice
model to those of a molecular dynamics simulation of slightly
polydisperse discs.

We characterize the statistical properties of the forces using the
force histogram and a two-point spatial correlation function of the
forces.  For the lattice model, in the granular limit the force
histogram has an exponential tail at large forces, while in the
elastic regime the force histogram is much narrower and has a form
that depends on the realization of disorder in the model.  The
behavior of the force histogram in the molecular dynamics simulations
as the pressure is increased is very similar to that displayed by the
lattice model.  In contrast, the spatial correlations evolve
qualitatively differently in the lattice model and in the molecular
dynamics simulations.  For the lattice model, in the granular limit
there are no in-plane stress-stress correlations, whereas in the
molecular dynamics simulation significant in-plane correlations
persist to the lowest pressures studied.
\end{abstract}
\pacs{45.70.Cc, 46.65.+g}

\widetext
\normalsize

\section{Introduction}
\label{sec:Intro}

Stress transmission in dry granular media is unusual because in these
materials no simple relation exists between stress and
strain\cite{sokolovskii65,nedderman92,edwards89,bouchaud95,
wittmer97}.  Physical ingredients that give rise to this are that
there are no tensile forces, that the particle deformations are very
small, and that the particles can rearrange~\cite{footnote1}.  Over
the last several years evidence has accumulated that force propagation
in dry granular media could be fundamentally different than in elastic
solids~\cite{edwards89,bouchaud95,wittmer96,wittmer97,cates98a,tkachenko99}.
Equations that have been proposed to describe stresses in lightly
loaded granular media have the property that specification of
boundary conditions at the top surface of the system is sufficient to
determine the stresses
throughout\cite{bouchaud95,wittmer96,wittmer97,cates98a,
tkachenko99,liu95,coppersmith96}, in marked contrast to the elliptic
equations of elasticity theory.

However, applying a large enough uniform pressure to a granular
material will cause it to exhibit an elastic linear response to a
small additional stress.  This is because uniform pressure both
inhibits rearrangements (because it suppresses Reynolds dilatancy) and
compresses the contacts, so that the non-tensile constraint on the
interparticle forces becomes irrelevant.  Thus, if stress propagation
in lightly loaded granular media is indeed substantially different
than in elastic media, then subjecting the material to high pressures
will change fundamentally the stress propagation characteristics.

This paper investigates theoretically the stress propagation in
granular materials as they are subjected to increasing pressures.  The
goals of this work are to understand the physical mechanisms governing
the evolution between granular and elastic behavior and to make
specific experimental predictions for the behavior of granular media
under increasing loads.

We study a two-dimensional model system and compare the results to
molecular dynamics (MD) simulations of two-dimensional systems of
slightly polydisperse discs.  Numerical studies of statistical models
of granular media, where geometrical complexity is modeled in terms of
uncorrelated random variables, are much faster and simpler than
molecular dynamics simulations.  Models of this type hold promise as a
means to obtaining insight into the physics underlying the force
propagation in granular materials.  Our model for the granular regime
is the two-dimensional scalar $q$-model~\cite{liu95, coppersmith96}.
Though the $q$-model has deficiencies~\cite{footnote2}, it is
attractive because of its simplicity and its prediction of an
exponential tail in the probability distribution of stress within a
packing agrees with experiments~\cite{miller96, baxter97, mueth98} and
with simulations~\cite{radjai96, radjai_unpub, radjai97, luding97,
clelland_unpub, thornton98}.  Our model for the elastic regime is a
network of springs with a regular topology, with disorder introduced
via randomly chosen spring
constants\cite{tang87,tang88,guyon90,sexton99}.  To model the
crossover between the two regimes, we exploit our observation that the
$q$-model can be written as a scalar elastic network subject to
certain constraints.  Enforcing these constraints to an increasing
degree, which causes the force propagation behavior to evolve from
that of an elastic system to that of the $q$-model, models the
crossover between elastic and granular behavior by a particulate
assemblage subjected to decreasing pressure.

We test the lattice model by comparing the results from the model to
those of our MD simulations of two-dimensional systems of slightly
polydisperse discs, focusing primarily on the probability distribution
of stresses and on the two-point stress-stress correlation functions.
The results of this investigation are mixed.  The crossover in the
force histogram between the elastic network and the $q$-model is
strikingly similar to the crossover observed in the molecular dynamics
simulations as the pressure on the system is decreased.  However, the
lattice model and the molecular dynamics simulation exhibit
qualitatively different trends in the behavior of the two-point
correlation functions of the stress.

The paper is organized as follows: Section~\ref{sec:Configuration}
defines the scalar networks that we investigate.
Section~\ref{sec:Methods} details the process of generation, solution,
and analysis of these networks and discusses the generation of the
molecular dynamics simulations of slightly polydisperse discs.
Section~\ref{sec:Results} reports the results of the force
distributions and spatial correlation functions for both the scalar
lattice model and the MD simulations.  Section~\ref{sec:Comparison}
compares the results of the scalar lattice model, the MD simulations,
and relevant experiments.  Section~\ref{sec:Discussion} summarizes and
interprets our results.  Appendix~\ref{app:Correlation} calculates a
finite-size correction to the in-plane stress-stress spatial
correlation function for the $q$-model that is relevant to the
interpretation of our numerical results.

\section{Scalar elastic networks and the $q$-model}
\label{sec:Configuration}
This section discusses the relationship between the $q$-model and the
elastic network studied in this paper.  Both models are scalar and are
defined on a two-dimensional lattice.  A scalar model is appropriate
for a spring network if either the the network is very highly
stretched~\cite{tang87, tang88, sahimi98}, or if the motions are
constrained so that displacements are unidirectional~\cite{sexton99}.
We consider the second situation and denote the direction along which
the motion occurs as $\hat{y}$, with positive $y$ pointing downwards.

\begin{figure}
\centering
\epsfig{file=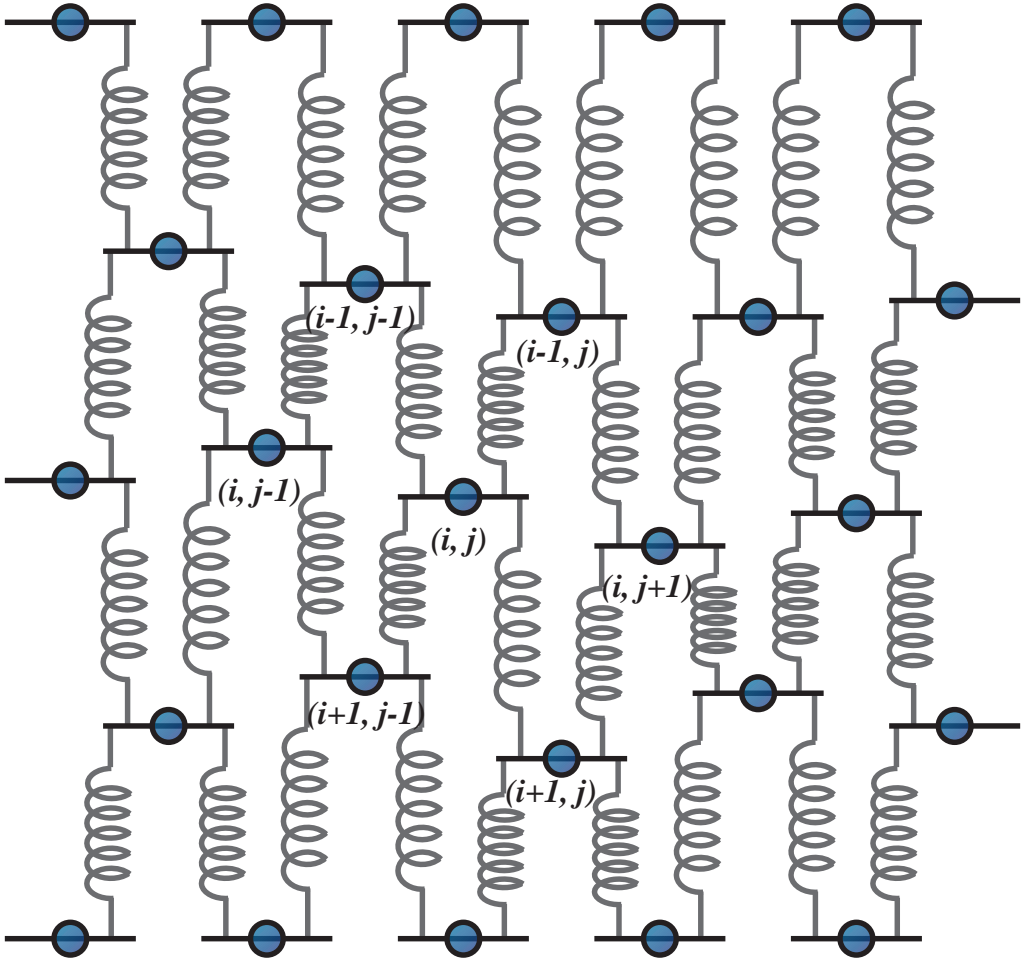, height=5in}
\\ \vspace{0.5in}
\caption{Elastic network considered in this paper.  Each node is
connected to two neighbors in the row above and to two neighbors in
the row below, with movement confined to the vertical direction.  A
node $(i,j)$ and its surrounding nodes have been labeled.  The system
is compressed by holding the top and bottom rows each at fixed
positions.  Disorder in the stress distribution is introduced by
variation of individual spring constants $k$.  In the elastic regime,
the $k$ values are chosen at random while in the granular regime, the
$k$'s are additionally constrained so that the strain in each row is
constant.}
\label{fig:lattice}
\end{figure}

Consider a network of nodes connected by springs on a diamond lattice
as shown in Fig.~\ref{fig:lattice}, where the motion of every node is
constrained to be along the vertical direction $\hat{y}$.  Each spring
has the same unstretched length, so that in the limit of zero load the
system forms a regular lattice.  The springs connecting the nodes have
spring constants that are chosen independently from a fixed
probability distribution.  Periodic boundary conditions are imposed in
the horizontal direction, and the locations of the nodes at the top
and bottom boundaries are fixed so that the vertical displacement of
all the nodes in these rows relative to the unloaded configuration are
identical.  We index the nodes so that a node in column $j$ in a row
$i$ with odd (even) $i$ lies along the same vertical line as the other
nodes in column $j$ in rows with odd (even) indices.

Let $y_{i,j}$ be the position of the node in row $i$ and column $j$
measured relative to its location in the absence of a load, and let
$k_{i,j}^l$ and $k_{i,j}^r$ be the spring constants of the springs
emanating downward from the node at row $i$ and column $j$.  Every
spring obeys Hooke's law, so that $f_{i,j}^l$ and $f_{i,j}^r$, the
forces exerted on node $(i,j)$ by the left and right springs below the
node, are $k_{i,j}^l(y_{i,j}-y_{i+1,j-1})$ and
$k_{i,j}^r(y_{i,j}-y_{i+1,j})$ for odd $i$
($k_{i,j}^l(y_{i,j}-y_{i+1,j})$ and $k_{i,j}^r(y_{i,j}-y_{i+1,j+1})$
for even $i$).  The system is then compressed by setting
$y_{1,j}=(L_{y} - 1) \Delta Y$ and $y_{L_{y},j}=0$ for all $j$, where
rows $1$ and $L_{y}$ are the top and bottom rows, respectively, and
$\Delta Y$ is the average strain.  We define $F_{i,j}$ to be the total
vertical force incident from above on node $(i,j)$, so that $F_{i,j} =
f_{i-1,j-1}^r + f_{i-1,j}^l$.  The forces and displacements are
determined by balancing the forces at every node, $F_{i,j} = f_{i,j}^l
+ f_{i,j}^r$, and requiring that each $y_{i,j}$ be well-defined.  This
latter condition can be written as $\mathbf{S} = -[d] \mathbf{Y}$;
here, $\mathbf{S}$ is the strain and $\mathbf{Y}$ is the displacement
field~\cite{bamberg90}.

In our spring networks, each spring constant has a value selected
independently at random from various probability distributions that
are described below.  We obtain the forces and strains along each link
of each network using the method outlined in Ref.~\cite{bamberg90}.

This scalar elastic model is equivalent to a resistor
network~\cite{tang87, tang88, sexton99}.  Forces and strains in the
elastic system correspond to currents and voltages, respectively, in the
resistor network.  The requirement that the vertical forces at each
node balance is equivalent to Kirchhoff's current law, while the
requirement that the position of each node is well-defined is
equivalent to Kirchhoff's voltage law.

\subsection{Comparison between the elastic model and the $q$-model.}
The force $F_{i,j}$ incident from above on node $(i,j)$ is transmitted
to the sites below in the two pieces $f_{i,j}^l$ and $f_{i,j}^r$.
Because of force balance, one can always write
\begin{equation}
\label{eq:division}
f_{i,j}^l = q_{i,j}F_{i,j}, ~~~f_{i,j}^r = (1-q_{i,j})F_{i,j}.
\end{equation}
In a $q$-model, the $q_{i,j}$ are random variables that are chosen
independently at every site.  In an elastic network,
Eq.~(\ref{eq:division}) still holds, but the $q_{i,j}$ are determined
by the configuration of random spring constants together with the
requirement that the displacement field be single-valued.  For spring
constants that are chosen independently, the force along any branch
will depend on the values of the spring constants throughout the
system.  Important consequences of this non-locality include the
presence of spatial correlations between the $q_{i,j}$'s and a
nontrivial relation between the distribution of spring constants and
the distribution of the $q$'s, including possibly the presence of
$q$'s that are negative, indicating the appearance of tensile forces
in the network.

A key observation underlying our work is that the $q$-model is
equivalent to an elastic network subject to the constraint that the
strain on every spring in each row is identical.  The strain need not
be constant from one row to the next, but it is simplest to consider
the case in which it is.  Let the amount of strain be $\Delta Y$.
Given the total force incident on node $(i,j)$ from above, $F_{i,j}$,
if one chooses the spring constants to be
\begin{equation}
\label{eq:q_Rvals}
k_{i,j}^l = \frac{q_{i,j} F_{i,j}}{\Delta Y} ,
~~~k_{i,j}^r = \frac{(1 - q_{i,j}) F_{i,j}}{\Delta Y},
\end{equation}
then the force exerted down the left link emanating from node $(i,j)$
is $k_{i,j}^l\Delta Y = q_{i,j} F_{i,j}$, and the force exerted down
the right link from node $i,j$ is $k_{i,j}^r\Delta Y = (1 - q_{i,j})
F_{i,j}$.  This force redistribution rule is exactly that of of the
$q$-model.  Given the set of $q_{i,j}$ values and the forces at each
node in the top row of the system, we can create an equivalent spring
network in a layer-by-layer manner.

We do not implement explicitly a no-tensile force constraint in our
networks, in contrast to the work of Refs. \cite{guyon90} and
\cite{sexton99}.  However, in the $q$-model limit, there are no
tensile forces.  Our molecular dynamics simulations of lightly loaded
material yield force distributions much closer to that of the
$q$-model than to those of the non-tensile elastic networks of
Ref.~\cite{sexton99}.

To study the crossover between elastic and $q$-model behavior, we
generate iteratively a sequence of networks that interpolate between
the elastic and $q$-model limits.  The procedure adjusts the spring
constants to make the strain in the system more uniform while keeping
the ratio of spring constants emanating from each node constant.  At
iteration $n$, the spring constants $k_{i,j}^{l}(n)$ and
$k_{i,j}^{r}(n)$ are set to
\begin{mathletters}
\begin{eqnarray}
\label{eq:q_Rvals_iterative}
k_{i,j}^l(n) &=&
        \frac{F_{i,j}(n-1)}{\Delta Y} q_{i,j} ~,\\
k_{i,j}^r(n) &=&
        \frac{F_{i,j}(n-1)}{\Delta Y} (1-q_{i,j}) ~,
\end{eqnarray}
\end{mathletters}
where $F_{i,j}(n-1)$ is the force through node $(i,j)$ at iteration
$n-1$.  The $q_{i,j}$ are kept fixed, and the iteration procedure is
started with $F_{i,j}(0) = 1$.

To characterize the crossover between elastic and $q$-model behavior
as the iteration proceeds, we need to quantify the degree to which the
constant-strain constraint is violated.  We use as our measure of the
spatial variation in the strain the dimensionless quantity
\begin{equation}
\label{eq:deviation}
\delta {\mathcal{S}}_{N} \equiv 
\frac{\displaystyle 
      \frac{1}{L_{x} (L_{y} - 1)} 
      \sum_{i=1}^{L_{y} - 1} \sum_{j=1}^{L_{x}} 
      \frac{1}{2} \left(
      (\delta Y_{i,j}^{l} - \overline{\delta{Y}})^{2} +
      (\delta Y_{i,j}^{r} - \overline{\delta{Y}})^{2}\right)}
     {\overline{\delta Y}^{2}}
\end{equation}
where $\delta Y_{i,j}^{l} = Y_{i,j} - Y_{i+1,j-1}$ and $\delta
Y_{i,j}^{r} = Y_{i,j} - Y_{i+1,j}$ for odd $i$ ($\delta Y_{i,j}^{l} =
Y_{i,j} - Y_{i+1,j}$ and $\delta Y_{i,j}^{r} = Y_{i,j} - Y_{i+1,j+1}$
for even $i$), and
\begin{equation}
\overline{\delta{Y}} =
      \frac{1}{L_{x} (L_{y} - 1)} 
      \sum_{i=1}^{L_{y} - 1} \sum_{j=1}^{L_{x}} 
          \frac{1}{2}(\delta Y_{i,j}^{l} + \delta Y_{i,j}^{r})
      = \Delta Y.
\end{equation}
Here, $L_{y}$ and $L_{x}$ are the number of rows and columns,
respectively.  In the elastic limit $\delta{\mathcal{S}}_{N} \approx
0.2$, and as discussed above, $\delta{\mathcal{S}}_{N}$ is zero for
the $q$-model.

\section{Methods}
\label{sec:Methods}

\subsection{Scalar lattice model}
We consider diamond-shaped lattices with springs on each link, as
shown in Fig.~\ref{fig:lattice}.  The positions of the top and bottom
node layers are fixed and periodic boundary conditions are imposed in
the transverse direction.  The forces along all the links depend on
the choice of spring constants, $\{k_{i,j}\}$, and are calculated
using the node-potential method described in Ref.~\cite{bamberg90}.
The overall strain for each network is scaled so that the average
vertical force through each node is normalized to unity,
\begin{equation}
\bar{F} \equiv \frac{1}{L_{x} L_{y}} 
          \sum_{i=1}^{L_{y}} \sum_{j=1}^{L_{x}} F_{i,j} = 1,
\end{equation}
where the sum is over the nodes in the network.

Networks of height $L_{y}=500$ are used, with analysis performed on
separate groups of layers to distinguish between edge and bulk
effects.  The widths $L_{x} = 16$ and $128$ are powers of 2 in order
to take advantage of FFT techniques in the calculation of spatial
correlation function values described below.  The number of
realizations averaged over varies from 10 to 50, depending on lattice
size and number of iterations.

For the elastic regime, we use four different distributions of spring
constants: uniform distribution of $k^{-1}$ for $k^{-1} \in (0,1)$,
gaussian distribution of $k^{-1}$ with the configuration average
$(\overline{k^{-1}}) = 1$ and standard deviation $\sigma_{k^{-1}} =
0.5$, uniform distribution of $k$ with $k \in (0,1)$, and gaussian
distribution of $k$ with $\bar{k} = 1$ and $\sigma_{k} = 0.5$.  We
construct networks with $L_{x} = 16$ and $128$ with 50 and 25
realizations, respectively.

For the $q$-model regime, a uniform distribution of $q$ with $q
\in (0,1)$ is used.  We implement the iterative scheme with networks
of size $L_{x} = 16$ and $128$ with 50 and 10 realizations,
respectively, for 100 iterations.

The local stress redistribution in a real granular material depends on
microscopic details such as particle shape, friction characteristics,
and preparation history.  Instead of attempting to model the local
force redistribution rules microscopically, our statistical models
treat them as random variables chosen from different probability
distributions.  Since these probability distributions are not known
{\em a priori}, we wish to identify and study properties that are not
sensitive to the choice of the probability distribution governing the
local force redistribution in the model.  We focus on $P(F)$, the
probability distribution of stresses at the nodes; $\tilde{P}(q)$,
the probability distribution of the redistribution fractions $q$; and
the spatial correlation functions of the force fluctuations about the
mean values~\cite{nicodemi98},
\begin{equation}
\label{eq:correlation}
C_{k}(j) = 
  \frac{1}{L_{y} - k} \sum_{l=1}^{L_{y} - k} \left(
  \frac{\displaystyle 
        \sum_{m=1}^{L_{x}} \delta F_{l,m} \delta F_{l+k,m+j}}
       {\displaystyle \sum_{m=1}^{L_{x}} \delta F_{l,m}^2}\right),
~~~\tilde{C}_{k}(j) = 
  \frac{1}{L_{y} - k} \sum_{l=1}^{L_{y} - k} \left(
  \frac{\displaystyle 
        \sum_{m=1}^{L_{x}} \delta q_{l,m} \delta q_{l+k,m+j}}
       {\displaystyle \sum_{m=1}^{L_{x}} \delta q_{l,m}^2}\right),
\end{equation}
where $\delta F_{i,j} = F_{i,j} - \bar{F}$ and $\delta q_{i,j} =
q_{i,j} - \bar{q}$; $\bar{F}$ is the average force and $\bar{q}$ is
the average $q$ value.  The indices $l$ and $m$ in
Eq.~\ref{eq:correlation} label layers and columns, respectively, while
$k$ and $j$ are the spatial separation in layers and columns.  These
correlation functions are normalized so that $C_{0}(0) = 1$ and
$\tilde{C}_{0}(0) = 1$.  Positive values (correlation) indicate a
tendency for nodes separated by $k$ rows vertically and $j$ columns
horizontally to be either both above or both below the mean, while
negative values (anti-correlation) indicates opposite behavior of one
above and one below the mean.

\subsection{Molecular dynamics simulations}

Here we discuss our molecular-dynamics (MD) simulations to used to
generate 2-D packings of discs.  Varying the ratio of external
load pressure to particle stiffness induces crossover between granular
and elastic behavior.  We calculate the probability distributions and
corresponding spatial correlation function values for forces and
redistribution fractions $q$ that are analogous to those in the scalar
model.

Our simulations employ a method similar to that used by Durian and
collaborators for sheared foams~\cite{durian95, durian97, langer97}, in
addition incorporating kinetic friction, contact damping, and particle
rotation, and using two different repulsive interparticle force laws
(linear and Hertzian).

\subsubsection{MD interaction rules}

The discs in our simulation are all of identical mass $m_{D} = 1$ and
interact via purely repulsive normal contact forces and kinetic
friction.  The interaction force between two discs whose centers are
at positions $\vec{r}$ and $\vec{r}_{j}$ with radii $a_{i}$ and
$a_{j}$ is non-zero only if their separation $\delta r_{i,j} \leq 0$,
where
\begin{equation}
\label{eq:overlap}
\delta r_{i,j} = \left|\vec{r}_{i} - \vec{r}_{j}\right| - (a_{i} + a_{j}).
\end{equation}
The normal contact force ${\mathcal{F}}_{i,j}$ is calculated from the
overlap $|\delta r_{i,j}|$.  We examine two force laws.  The first is
a linear force law based on a spring-like restoring force that yields
\begin{equation}
\label{eq:linear_contact}
{\mathcal{F}}_{i,j} = K_{i,j} \left|\delta r_{i,j}\right|,
\end{equation}
with $K_{i,j} = (1/K_{i} + 1/K_{j})^{-1}$, where $K_{d}$ is the spring
constant for disc $d$.  The second is a non-linear force law based on
Hertzian contacts between spheres\cite{landau_elastic},
\begin{equation}
\label{eq:hertzian_contact}
{\mathcal{F}}^{[\mathrm{HC}]}_{i,j} = 
	D^{-1} (\frac{1}{a_{i}} + \frac{1}{a_{j}})^{-\frac{1}{2}} 
        \left|\delta r_{ij}\right|^{\frac{3}{2}},
\end{equation}
where $D = \frac{3}{2} \left((1 - \sigma^{2}) / E\right)$, with
$\sigma$ and $E$ being the material properties Poisson's ratio and
Young's modulus, respectively.  For both force laws, the forces are
directed so as to separate the overlapping discs.  To calculate forces
generated by interactions with walls, we assume the walls to be discs
of infinite radius.

Kinetic friction is incorporated into the disc interactions although
static friction is not.  The introduction of frictional forces causes
the discs to rotate; however, the frictional force is zero at
mechanical equilibrium.  The kinetic friction force $f_{i,j}$ for
contact between discs $i$ and $j$ is
\begin{equation}
f_{i,j} = \mu {\mathcal{F}}_{i,j},
\end{equation}
where $\mu$ is the coefficient of kinetic friction, and is directed
opposite to the contact point velocity $\vec{v}^{\mathrm{cp}}_{i,j}$.
For disc $i$, this velocity $\vec{v}^{\mathrm{cp}}_{i,j}$ is related
to disc velocities $\vec{v}_{i}$ and $\vec{v}_{j}$, the angular
velocities $\omega_{i}$ and $\omega_{j}$, and directional vector
$\hat{r} = (\vec{r}_{i} - \vec{r}_{j}) / |\vec{r}_{i} - \vec{r}_{j}|$
by
\begin{equation}
\vec{v}^{\mathrm{cp}}_{i,j} = 
  \vec{v}_{i,j} - \left(\vec{v}_{i,j} \cdot \hat{r}\right) \hat{r} 
   + \left(a_{i} \vec{\omega}_{i} + a_{j} \vec{\omega}_{j}\right)
     \times \hat{r},
\end{equation}
where $\vec{v}_{i,j} = \vec{v}_{i} - \vec{v}_{j}$.  

Damping during contact between discs $i$ and $j$ is used as an
additional means of dissipating kinetic energy.  It is generated by
applying to disc $i$ a force ${\mathcal{F}}_{\mathrm{D}}$ and torque
$\Gamma_{\mathrm{D}}$ given by
\begin{mathletters}
\begin{eqnarray}
{\mathcal{F}}_{\mathrm{D}} & = & -\lambda_{\mathrm{trans}} v'_{i},\\
\Gamma_{\mathrm{D}} & = & -\lambda_{\mathrm{ang}} \omega'_{i},
\end{eqnarray}
\end{mathletters}
where $v_{i}'$ is the translational velocity of disc $i$ relative to
the interaction center of mass for the two discs $i$ and $j$ that are
in contact and $\omega'_{i}$ is its angular velocity relative to the
total angular momentum of the disc pair.  $\lambda_{\mathrm{trans}}$
and $\lambda_{\mathrm{ang}}$ are damping constants.  This process
conserves both translational and angular momentum.  Energy is directly
removed from the system as opposed to being converted between
translational and rotational motion.

The bottom and top walls have mass $m_{W}$ and are constrained to move
only vertically.  An inward force of magnitude $F_{\mathrm{wall}}$ is
applied to each wall in order to compress the system.  Damping of the
wall motion suppresses volume oscillations and serves as the primary
means of energy dissipation.  The damping force
${\mathcal{F}}_{\mathrm{WD}}$ on a wall is
\begin{equation}
{\mathcal{F}}_{\mathrm{WD}} = -\lambda_{W} v_{W},
\end{equation}
where $v_{W}$ is the velocity of the wall and $\lambda_{W}$ is the
wall damping constant.

\subsubsection{MD implementation}

Ensembles of systems of $N=1024$ discs of average radius $a_{D}$ are
generated by starting with triangular array of $\sqrt{N}$ rows and
$\sqrt{N}$ discs per row placed in a horizontally periodic system with
both height and width $L = 2.273 a_{D} \sqrt{N}$.  For the data shown
here, discs are placed in the system at positions $(L(n_{x} +
0.05)/\sqrt{N}, L(n_{y} + 0.5)/\sqrt{N})$ for odd $n_{y}$ and
$(L(n_{x} + 0.55)/\sqrt{N},L(n_{y} + 0.5)/\sqrt{N})$ for even $n_{y}$,
with indices $n_{x}$ and $n_{y}$ running from 0 to $\sqrt{N} - 1$.  In
practice, discs with gaussian distributed polydispersity of
$\sigma_{a} = 0.1 a_{D}$ placed on this triangular array do not
overlap.  The results obtained are not sensitive to initial disc
placement.  The system is then compressed by the application of an
inward force on the top and bottom walls.  All discs have the same
spring constant $K_{d} \equiv K = 1$.  The coefficient describing wall
damping is set to $\lambda_{W} / m_{W} = 1$.  Damping coefficients
for translational and angular motion for disc contacts are set to
$\lambda_{\mathrm{trans}} / m_{D} = 1$ and $\lambda_{\mathrm{ang}} /
I_{D} = 4.1$, where $I_{D} =
\frac{1}{2} m_{D} a_{D}^{2}$ is the moment of inertia for a disc with
radius $a_{D}$.  The coefficient of kinetic friction is set to $\mu =
0.2$ for both disc-disc and disc-wall contacts.  Comparisons with
samples produced without disc-contact damping or kinetic friction
revealed no measurable differences in force probability distributions
or in the two-point force correlation function.  Incorporating
additional energy-dissipation mechanisms allows systems to reach
mechanical equilibrium more rapidly.  The end time for each
compression stage is chosen so that the average residual kinetic
energy for each disc is equivalent to translational movement of
approximately or less than $0.01 a_{D}$ in unit time.  Because of the
increased external energy input, systems at higher compressions are
allowed a less restrictive limit of approximately $0.05 a_{D}$.
Visual inspection of final configurations do not reveal significant
fluctuations in time in contact network topology or force magnitude in
load-bearing structures.  Comparisons with test systems with longer
run times also do not show any significant quantitative differences.

\begin{figure}
\centering
\begin{tabular}{cc}
\subfigure[$\Pi = 0.1 \Pi_{0}$]{\epsfig{file=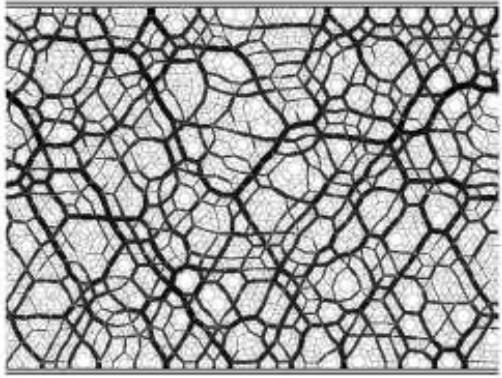, width=3.25in}}
&
\subfigure[$\Pi = \Pi_{0}$]{\epsfig{file=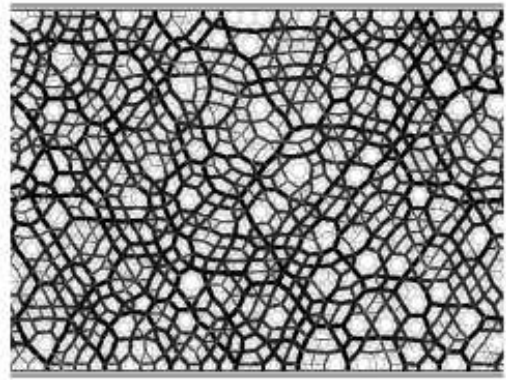, width=3.25in}}
\\
\subfigure[$\Pi = 10 \Pi_{0}$]{\epsfig{file=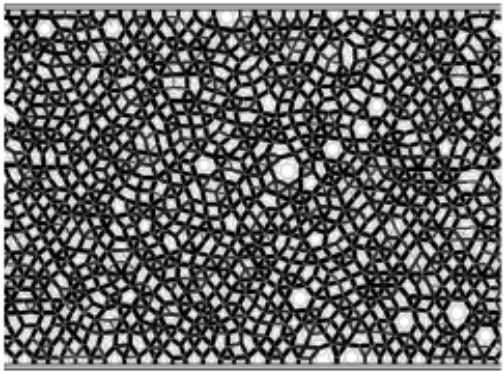, width=3.25in}}
&
\subfigure[$\Pi = 50 \Pi_{0}$]{\epsfig{file=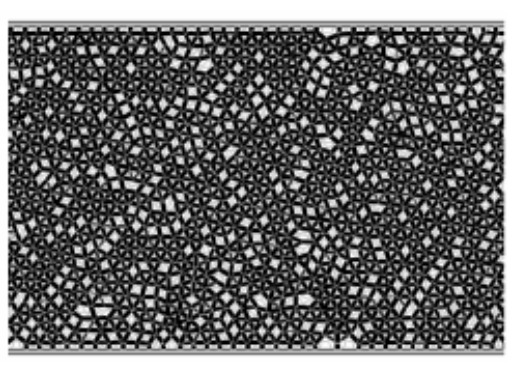, width=3.25in}}
\end{tabular}
\caption{Contact networks of a sample MD-generated packing of 1024
discs for different values of applied external pressure $\Pi$.  The
reference pressure $\Pi_{0}$ is such that the average fractional
change in particle radius at a contact is $6.25 \times 10^{-4}$.  In
the transition from granular to elastic behavior, the number of
contacts in the system increases and the magnitudes of the contact
forces become more homogeneous.  While the width of contacts shown is
a proportional to the force magnitude, they have been rescaled for
each pressure so direct comparisons between subfigures is not
possible.}
\label{fig:md_sample}
\end{figure}

For a system of fixed size $L$, the typical compression of the system
can be controlled through variations in the disc spring constant
$K$ or applied external force $F_{\mathrm{wall}}$.  Typical
relative particle deformations $\delta {\mathcal{R}}$ is given by
\begin{equation}
\label{eq:md_deviation}
  \delta {\mathcal{R}} \equiv
  \frac{1}{N_{C}} \sum_{(i,j)} \frac{|\delta r_{i,j}|}{(a_{i} + a_{j})} =
  \frac{1}{N_{C}} \sum_{(i,j)} 
     \frac{{\mathcal{F}}_{i,j}}{K_{i,j}}(a_{i} + a_{j})^{-1}
  \approx \frac{F_{\mathrm{wall}} \left( \frac{L}{a_{D}}
 	   \right)^{-1}}{\frac{K}{2}} (2 a_{D})^{-1} =
  \frac{F_{\mathrm{wall}}}{L K} =
  \frac{\Pi}{K},
\end{equation}
where $N_{C}$ is the total number of contacts, the sums are over pairs
of discs $i$ and $j$ in contact, and $\Pi \equiv F_{\mathrm{wall}}/L$
is the external pressure.  This estimate is approximate due to
geometric factors and distributional fluctuations; however, the
scaling of deformations to $\Pi/K$ should hold generally.  In our
simulations, the disc spring constant $K$ is held fixed and the
pressure $\Pi$ is varied to induce crossover between granular and
elastic behavior.  We define the reference pressure $\Pi = \Pi_{0}$
such that the relative particle deformation $\delta {\mathcal{R}}
\approx 6.25 \times 10^{-4}$.  The reference compression pressure
$\Pi_{0}$ yields a force histogram typical of the granular range, as
discussed below in Sec.~\ref{sec:Results}.  After the initial
compression with $\Pi = \Pi_{0}$, the applied pressure is increased in
stages to $\Pi = 100 \Pi_{0}$, at which $\delta {\mathcal{R}} \approx
0.01$.  We also decrease the pressure from the initial $\Pi = \Pi_{0}$
configuration down to $\Pi = 0.01 \Pi_{0}$ ($\delta {\mathcal{R}}
\approx 10^{-6}$) in order to approach the zero-deformation limit.
Fig.~\ref{fig:md_sample} shows a sample MD system subjected to the
pressures $0.1 \Pi_{0}$, $\Pi_{0}$, $10 \Pi_{0}$, and $50 \Pi_{0}$.

For spheres with Hertzian contacts (using
Eq.~\ref{eq:hertzian_contact}), the deformation can be approximated by
$\delta {\mathcal{R}}^{\mathrm{[HC]}} \approx \left( \frac{\Pi D}{2
a_{D}} \right)^{\frac{2}{3}}$.  For our simulations $D$ is chosen to
yield deformations of the same order of magnitude as the linear
contacts at the compression $\Pi = \Pi_{0}$.  The pressures
studied are the same as for the linear spring contact systems.

\section{Results}
\label{sec:Results}

\subsection{Scalar lattice model}
Here we present our results for the scalar lattice models.  We study
how the probability distribution of total vertical force $F$ incident
on a node from above $P(F)$ and the two-point force correlation
function $C_{k}(j)$ characterize the behavior in scalar elastic
lattice networks in which the constant-strain constraint is enforced
to varying degrees.  In the $q$-model both $P(F)$ and in-plane
force-force correlation function $C_{0}(j)$ exhibit robust behavior
for generic choices of probability distributions of $q$'s.  We
investigate the degree to which these quantities depend on the choice
of spring constant distributions in the elastic networks, and discuss
the crossover of $P(F)$ and $C_{0}(j)$ between the elastic and
$q$-model behavior as the constant-strain constraint is implemented
with increasing accuracy.

\subsubsection{Results for the $q$-model}
In the $q$-model, the force histogram $P(F)$ decays exponentially at
large forces\cite{coppersmith96} and $C_{0}(j)$ is zero for non-zero
$j$~\cite{coppersmith96,krug99,majumdar99}.  These properties hold for a
wide variety of choices of the distribution of $q$-values.

Our results for the crossover from elastic to $q$-model behavior are
obtained for the specific choice that the $q$'s are uniformly
distributed in $[0,1]$.
A two-dimensional $q$-model with this distribution
of $q$'s yields\cite{coppersmith96}
\begin{equation}
P(F) = 4 F e^{-2 F}.
\end{equation}
For a system of infinite lateral extent, the in-plane force-force
correlation function $C_{0}(j)=\delta_{j0}$ where $\delta_{j0}$ is the
Kronecker delta function \cite{coppersmith96}.  For a system of finite
width $L_{x}$, force correlations must arise because all forces are
positive and the total force through a layer is fixed.  As discussed
in Appendix \ref{app:Correlation}, assuming that this mechanism is the only
one giving rise to correlations, one obtains that a 2-D system of
lateral extent $L_{x}$ has $C_{0}(j)$ given by
\begin{equation}
C_{0}(j \neq 0) = -(L_{x} - 1)^{-1}.  
\end{equation} 
This form for $C_{0}(j)$ agrees with our numerical results for the
$q$-model on lattices of finite width.

\subsubsection{Elastic networks}
\begin{figure}
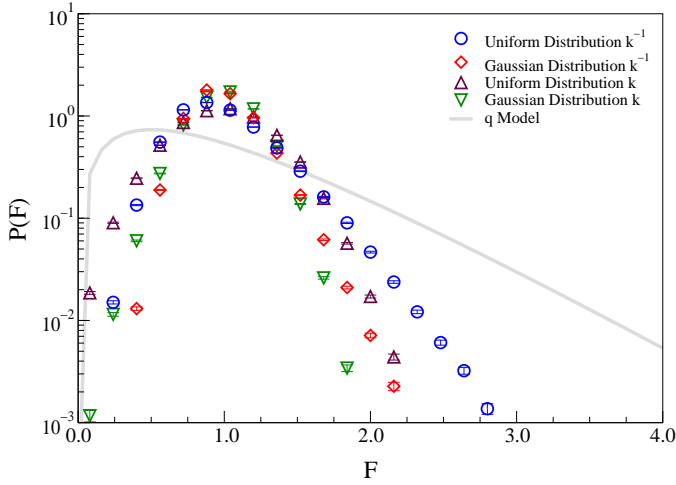
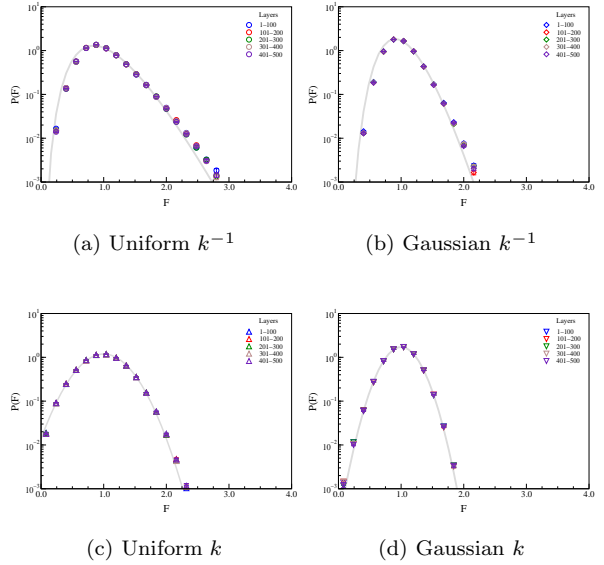

\centering
\begin{tabular}[t]{cc}
\epsfig{file=scalar_pF_128c.eps, width=3.5in} &
\begin{minipage}[t]{3.25in}
\begin{tabular}[b]{cc}
\subfigure[Uniform $k^{-1}$]{\epsfig{file=uni_kinv_pF_128c.eps, width=1.5in}} &
\subfigure[Gaussian $k^{-1}$]{\epsfig{file=gauss_kinv_pF_128c.eps, width=1.5in}} \\
\subfigure[Uniform $k$]{\epsfig{file=uni_k_pF_128c.eps, width=1.5in}} &
\subfigure[Gaussian $k$]{\epsfig{file=gauss_k_pF_128c.eps, width=1.5in}}
\end{tabular}
\end{minipage}
\end{tabular}
\caption{Force probability distribution $P(F)$ for 128-column random
spring configurations. The form of $P(F)$ in the elastic regime
depends on the distribution of spring constant values $k$.  However,
these $P(F)$ for all distributions of $k$ occupy a narrow range in
comparison with the $q$-model granular regime result shown by the
solid gray line.  (a)-(d) We fit the functional form $P(F) \propto
F^{A} e^{-B F}$ to the random $k^{-1}$ distributions with
$A=6.67$ and $B=7.95$ for the uniform distribution and
$A=14.70$ and $B=16.21$ for the gaussian distribution.  For
the random $k$ distributions, we fit $P(F) \propto
e^{(F-1)^{2}/S^{2}}$ with $S=0.47$ for the uniform distribution and
$S=0.32$ for the gaussian distribution.  No differences are seen
between layer groups near the edge or in the bulk of the systems.}
\label{fig:scalar_pF_128c.dists}
\end{figure}

For elastic networks with different distributions of spring constants,
the probability distribution of vertical force $P(F)$, shown in
Figs.~\ref{fig:scalar_pF_128c.dists}, is narrower than that of the
$q$-model.  Its functional form depends on the choice of spring
constant distribution.  Choosing the spring constants $k$ from a
distribution either uniform in $k$ or gaussian in $k$ yields $P(F)$'s
that are roughly gaussian while the $P(F)$'s for networks for
distributions uniformly distributed or gaussian in $k^{-1}$ display a
tail at large $F$ that is consistent with an exponential decay.
Networks with gaussian distributed $k$ or $k^{-1}$ exhibit narrower
$P(F)$'s than their counterparts with uniformly distributed $k$ or
$k^{-1}$.

\begin{figure}
\centering
\epsfig{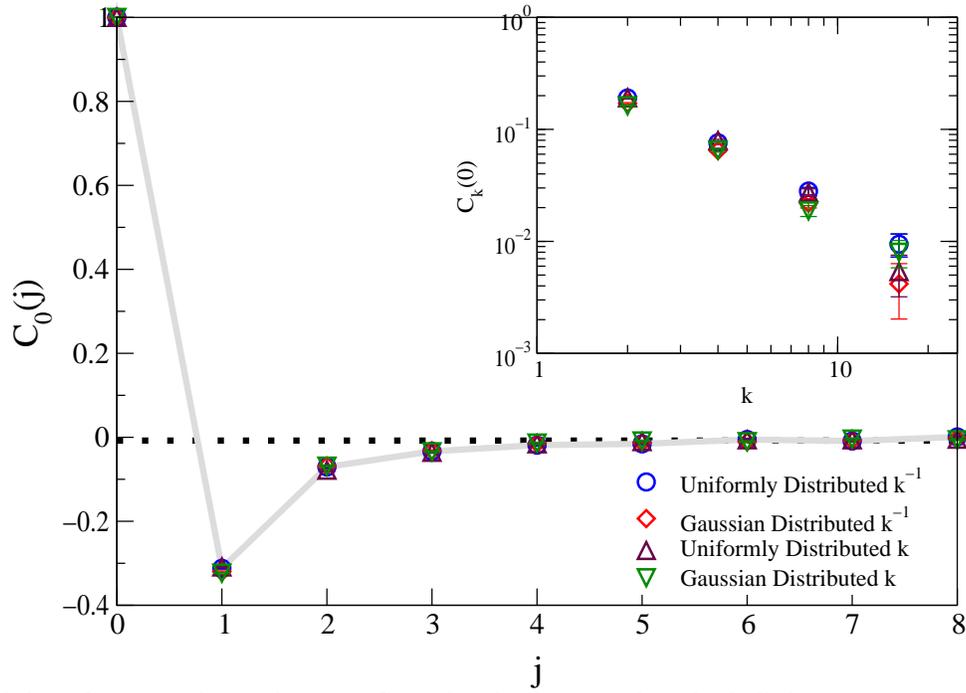}
\caption{Spatial force-force correlation function $C_{k}(j)$ for the
forces within the bulk layer grouping (layers 201-300) in 128-column
elastic-regime scalar networks.  In-plane correlations $C_{0}(j)$ are
shown in the main plot while the vertical correlation peak decay given
by $C_{k}(0)$ is shown in the inset.  An observed nearest-neighbor
in-plane anti-correlation appears to be robust with respect to
variations in spring constant distributions.  The vertical correlation
is similarly robust.}
\label{fig:scalar_cF_128c}
\end{figure}

In contrast to the behavior of the force probability distribution 
$P(F)$, the force-force correlation function values $C_{k}(j)$ are 
quantitatively indistinguishable for all the distributions of spring 
constants that we examined, as shown in Fig.~\ref{fig:scalar_cF_128c}.  
For $C_{0}(j)$, the force-force correlation function within the same
layer, we see a strong anti-correlation for $j = 1$ of magnitude $\sim
0.30$ that decays within $j \leq 8$.  For vertical separation $k > 0$, 
we see a simultaneous reduction in peak magnitude (at $j = 0$) and 
broadening of peak width but with the anti-correlation signature and 
decay joining the curve laid out by $k = 0$.

\begin{figure}
\centering
\epsfig{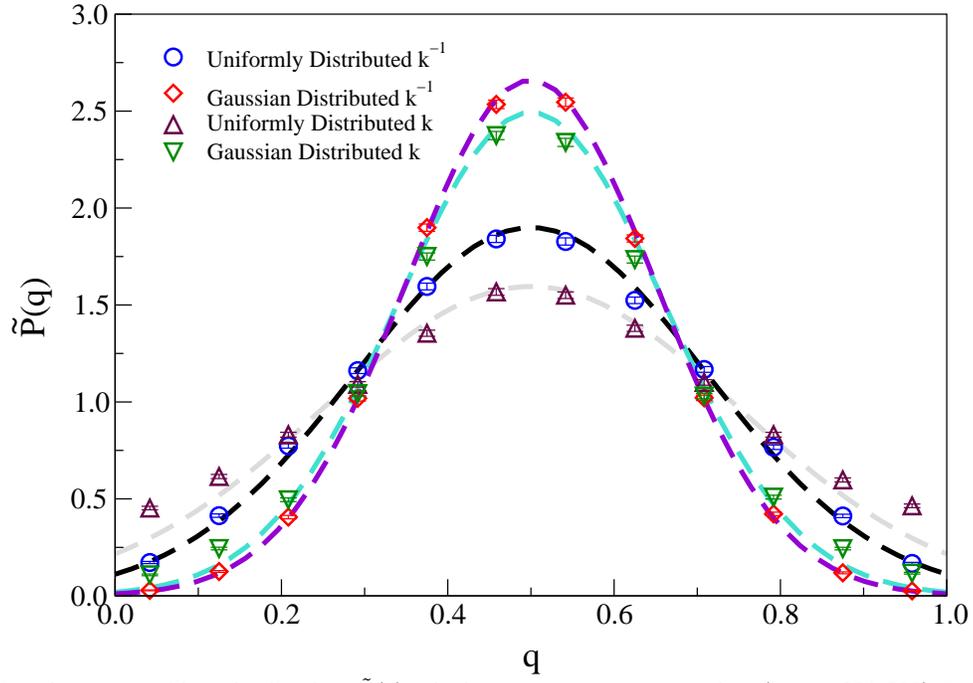}
\caption{Force fraction probability distribution $\tilde{P}(q)$ within
the bulk layer grouping (layers 201-300) for 128-column elastic-regime
scalar networks.  The fits are gaussians peaked at $q=0.5$ with the
width being dependent on the spring constant distributions.  Gaussian
distributed $k^{-1}$ and $k$ configurations are narrower, with widths
$\sigma \approx 0.16$ and $0.15$, respectively, than their uniformly
distributed counterparts (random $k$, $\sigma \approx 0.25$ and random
$k^{-1}$, $\sigma \approx 0.21$).}
\label{fig:scalar_pq_128c}
\end{figure}

\begin{figure}
\centering
\epsfig{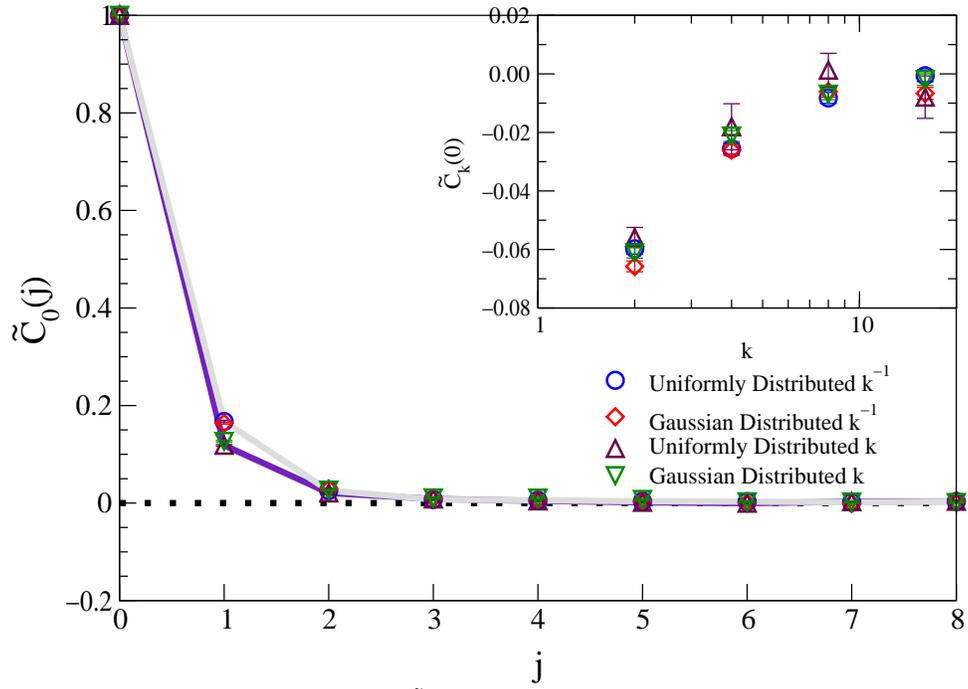}
\caption{The two-point spatial correlation function $\tilde{C}_{k}(j)$
for the force fraction $q$ within the bulk layer grouping (layers
201-300) in 128-column elastic-regime scalar networks.  In-plane
correlations $\tilde{C}_{0}(j)$ are shown in the main plot with
vertical correlations $\tilde{C}_{k}(0)$ in the inset.  As with the
force-force correlations, varying the spring constant distribution has
minimal effect on these correlations.}
\label{fig:scalar_cq_128c}
\end{figure}

\begin{figure}
\centering
\epsfig{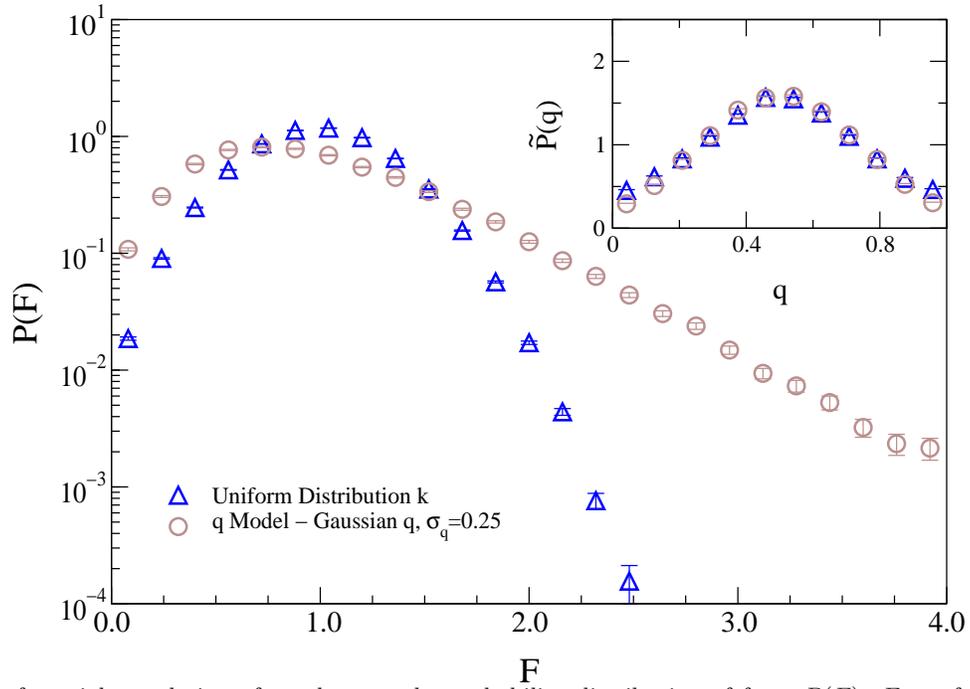}
\caption{Effect of spatial correlation of $q$ values on the
probability distribution of force $P(F)$.  Force fraction probability
distributions $\tilde{P}(q)$ of the uniform distribution of $k$
elastic network and a $q$-model system generated by choosing $q$
values from a gaussian distribution centered at $q=0.5$ with width
$\sigma_{q} = 0.25$ is shown in the inset.  While the two
$\tilde{P}(q)$ distributions appear nearly identical, the spatial
correlations in the elastic network yield a functionally different
form for the probability distribution of forces $P(F)$ than that of
the uncorrelated $q$-model system.}
\label{fig:qm_no_correlation}
\end{figure}

The probability distributions of redistribution fraction $q$,
$\tilde{P}(q)$, shown in Fig.~\ref{fig:scalar_pq_128c}, are 
roughly gaussian and peaked at $q = \bar{q} \equiv 0.5$ for all
distributions of spring constants examined.  The widths of the
$\tilde{P}(q)$ depend on the choice of distribution of spring
constants, with the gaussian distributed $k$ and $k^{-1}$ once again
being narrower (standard deviation $\sigma_{q} \approx 0.16$ and
$0.15$, respectively) than their uniformly distributed counterparts
($\sigma_{q} \approx 0.25$ for random $k$ and $\sigma_{q} \approx
0.21$ for random $k^{-1}$).  All of the elastic networks display
significant correlations between $q$'s at different nodes as
demonstrated in Fig.~\ref{fig:scalar_cq_128c}, which shows the
correlation function $\tilde{C}_{k}(j)$ for all the random
distributions.  The correlations between $q$'s are an important factor
in determining the statistical distribution of the forces in these
systems; Fig.~\ref{fig:qm_no_correlation} shows that a $q$-model
system with the same $\tilde{P}(q)$ as an elastic network with a
uniform distribution of $k$ but with no correlations between the $q$'s
yields a $P(F)$ markedly different from the elastic network.

No differences between the bulk (layers 201-300) and edge (layers
1-100 and 401-500) sections are detected in the distributions $P(F)$ and
$\tilde{P}(q)$ or the correlation functions $C_{k}(j)$ and
$\tilde{C}_{k}(j)$.  The results for lattices with $L_x=128$ are the
same within statistical errors to the results from $L_{x}=16$
lattices.

In the elastic networks, forces in less than 1\% of the branches are
tensile, and no node in any of the networks is subject to a tensile
net force.  Our results for $P(F)$ for uniformly distributed $k$'s are
very similar to those reported in Sexton \textit{et
al.}~\cite{sexton99}, where a non-tensile force constraint is
enforced.

\subsubsection{Iterated networks-- the $q$-model limit}

We now discuss the networks generated by our iterative algorithm for
converting an elastic network to a $q$-model system.  First, we
verify that the generated networks do eventually converge to the
$q$-model.  After 100 iterations, the forces along the links of the
iterated spring network are identical to those of the corresponding
$q$-model to within $10^{-4}$.  

A subtle point in the method is that our iterative scheme yields a
configuration in which the forces at the top and bottom boundaries of
the iterated network may have nonzero spatial correlations, as the
initial iteration $n=0$ system is elastic.  As one proceeds away from
the top and bottom boundaries, these correlations decay via a
diffusive process that takes on the order of $L_{x}^{2}$
layers~\cite{coppersmith96}.  Thus, forces at different sites in the
same layer are effectively uncorrelated only for systems with large
aspect ratios.  This result is consistent with our numerical
observation that in fully iterated systems, correlations between
forces at different sites in the same layer are present throughout the
$L_{x}=128$ systems, while they are only present in the top- and
bottom-most 200 layers of $L_{x}=16$ systems.

\subsubsection{Iterated networks-- crossover between elastic and $q$-model 
               behavior}

In the iterated networks the target values of the $q_{i,j}$ are fixed
at the outset of iteration procedure.  The initial state (zeroth
iteration) is an elastic network with spring constants given by
$k_{i,j}^{l} = q_{i,j}/\Delta Y$ and $k_{i,j}^{r} = (1-q_{i,j})/\Delta
Y$.  Therefore, the initial probability of node forces $P(F)$ and the
spatial correlation function $C_{0}(j)$ are those of elastic networks
with spring constants chosen from a uniform distribution of $k$.  The
realized $q$ distribution $\tilde{P}(q)$ (as opposed to the
distribution of the target $q$ values) is peaked at $q = 0.5$ and its
spatial correlation function $\tilde{C}_{k}(j)$ reveals slight
nearest-neighbor correlations for $k=0$ and anti-correlations at $j =
0$ for $k > 0$, once again matching elastic-regime behavior.

\begin{figure}
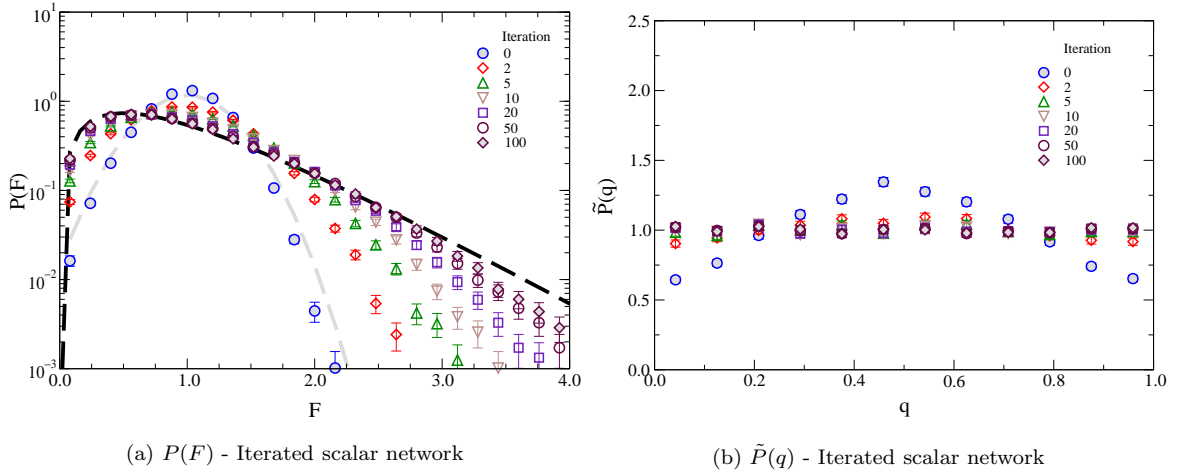

\centering
\begin{tabular}{cc}
\subfigure[$P(F)$ - Iterated scalar network]
{\epsfig{file=iter_pF_16c.eps, width=3in}} &
\subfigure[$\tilde{P}(q)$ - Iterated scalar network]
{\epsfig{file=iter_pq_16c.eps, width=3in}}
\end{tabular}
\caption{(a) Force probability distribution $P(F)$ at various stages
of iteration for the bottom layer grouping (401-500) of an ensemble of
16-column scalar networks.  Initially $P(F)$ is similar to the
distribution for uniformly distributed $k$ systems, shown by the grey
dashed line.  The distribution broadens with increasing iterations with
small $F$ agreement with granular $q$-model systems being achieved on
order of 20 iterations.  Further iterations are necessary to approach
agreement for large values of $F$.  The $P(F)$ distribution for
$q$-model systems is shown by the black dashed line.  (b) Force
fraction probability distribution $\tilde{P}(q)$ at corresponding
stages of iteration.  The distribution of $q$ values approaches the
expected uniform distribution within the first 10 iterations.}
\label{fig:iter_pFq_16c}
\end{figure}

The probability distribution of node forces, $P(F)$, is shown in
Fig.~\ref{fig:iter_pFq_16c}(a) for different values of iteration
number $n$.  As the number of iterations is increased, the $P(F)$
develops an exponential tail at large forces.
Fig.~\ref{fig:iter_pFq_16c}(b) shows the probability distribution of
the $q$'s, $\tilde{P}(q)$, versus the number of iterations.
$\tilde{P}(q)$ approaches the target form of a uniform distribution
after roughly 10 iterations.

\begin{figure}
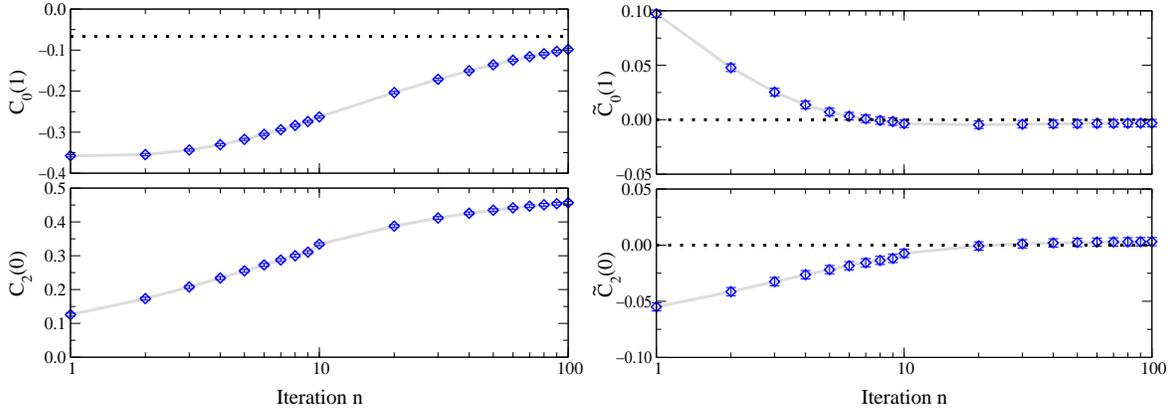

\centering
\begin{tabular}{cc}
\subfigure[Force-force correlations - Iterated scalar network]
{\epsfig{file=iter_cF_16c.eps, width=3in}} &
\subfigure[$q$-$q$ correlations - Iterated scalar network]
{\epsfig{file=iter_cq_16c.eps, width=3in}}
\end{tabular}
\caption{(a) Nearest-neighbor in-plane and vertical force-force
correlation values $C_{0}(1)$ and $C_{2}(0)$ at various stages of
iteration for an ensemble of 16-column scalar networks.  The observed
anti-correlation in nearest-neighbor forces decreases in magnitude
as the number of iterations increases and appears to approach
asymptotically the expected zero-correlation value.  The vertical
correlation increases in magnitude with increasing iterations,
indicating a stronger preference for the formation of force vertical
channels.  (b) Nearest-neighbor in-plane and vertical $q$-$q$
correlation values $\tilde{C}_{0}(1)$ and $\tilde{C}_{2}(0)$.  The
spatial correlations for the force fraction $q$ decrease rapidly in
magnitude as their values approach the uncorrelated target $q$
distribution.}
\label{fig:iter_cFq_16c}
\end{figure}

Fig.~\ref{fig:iter_cFq_16c}(a) shows our results for nearest-neighbor
in-plane and vertical force-force correlation function values
$C_{0}(1)$ and $C_{2}(0)$ as the number of iterations $n$ is
increased.  Fig.~\ref{fig:iter_cFq_16c}(b) shows the corresponding
force-fraction $q$-$q$ correlations $\tilde{C}_{0}(1)$ and
$\tilde{C}_{2}(0)$.  While only about 10 iterations are necessary
before the nearest-neighbor spatial correlations between $q$ values go
to zero, in-plane force-force correlations are still present after 100
iterations although much reduced in magnitude from the initial elastic
(iteration $n = 0$) value and approaching asymptotically the expected
zero-correlation value.

\begin{figure}
\centering
\epsfig{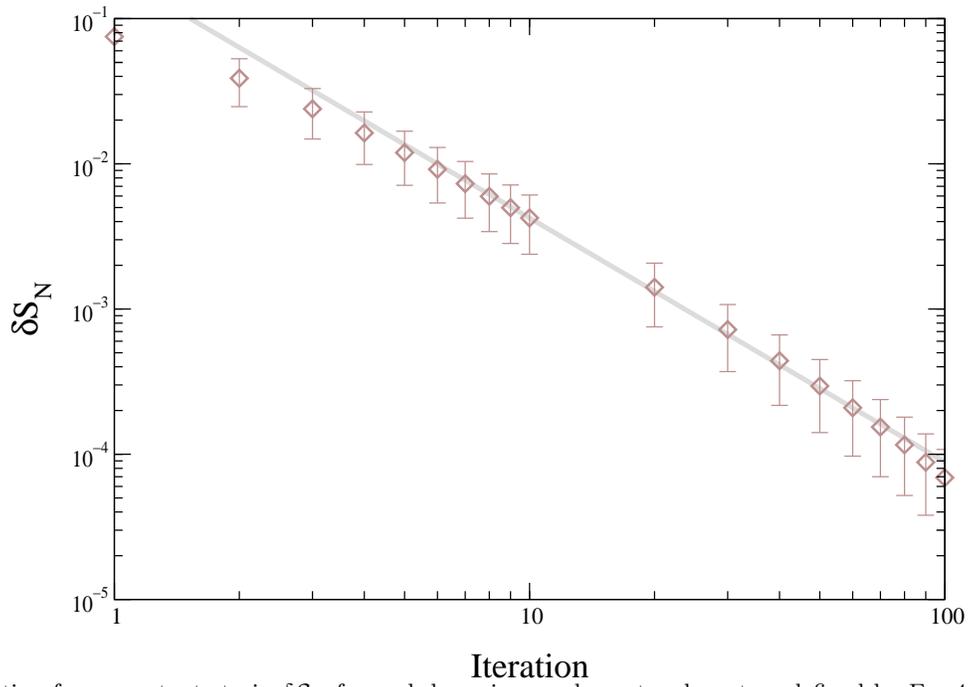}
\caption{Deviation from constant strain $\delta{\mathcal{S}}_{N}$ for
each layer in a scalar network system defined by
Eq.~\ref{eq:deviation}.  The solid line is a power law fit of form
$\delta {\mathcal{S}}_{N} \propto n^{-\alpha}$ with $\alpha = -1.68
\pm 0.02$.  Increasing iterations confirm the approach to the constant
strain limit, which is equivalent to the $q$ model.}
\label{fig:iter_dS_16c}
\end{figure}

The quantity $\delta{\mathcal{S}}_{N}$ that we use to characterize the
crossover between elastic and $q$-model behavior is defined in
Sec.~\ref{sec:Configuration}.  As Fig.~\ref{fig:iter_dS_16c}
demonstrates, we observe $\delta{\mathcal{S}}_{N}$ decreasing as the
number of iterations $n$ is increased according to a power law.  A fit
that assumes the dependence is of the form $\delta{\mathcal{S}}_{N}
\propto n^{\alpha}$ yields $\alpha = -1.68 \pm 0.02$.

\subsection{Results of molecular dynamics simulations}

\begin{figure}
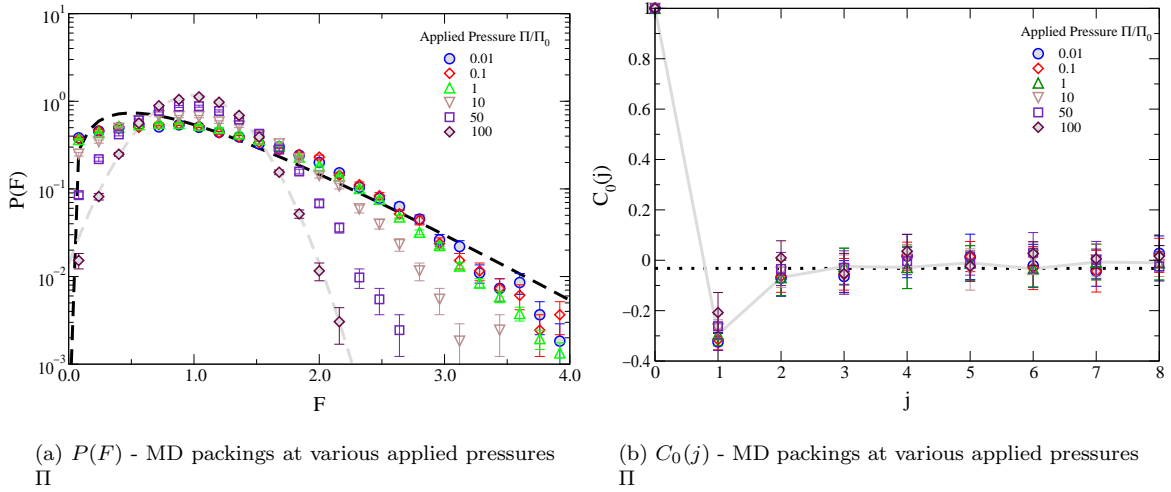

\centering
\begin{tabular}{cc}
\subfigure[$P(F)$ - MD packings at various applied pressures $\Pi$]
{\epsfig{file=md_pF.eps, width=3in}} &
\subfigure[$C_{0}(j)$ - MD packings at various applied pressures $\Pi$]
{\epsfig{file=md_cF.eps, width=3in}}
\end{tabular}
\caption{(a) Progression of vertical force probability distribution
$P(F)$ between elastic and granular regimes in MD-generated packings.
At high applied pressure $\Pi$, $P(F)$ has a gaussian functional form,
as shown by the gray dashed line, similar to the elastic behavior of
the scalar network.  At low pressure, $P(F)$ has widened and displays
an exponential tail.  For comparison, $P(F)$ for the $q$-model is
shown by the black dashed line.  The apparent transition to the
granular regime appears to occur by roughly $\Pi / \Pi_{0} = 1$. (b)
In-plane force correlation function value $C_{0}(j)$ for various
applied pressures.  We define discs to be in-plane within a tolerance
of $\pm0.10 a_{D}$ with spacings $j$ given in units of average disc
diameter $2 a_{D}$.  Unlike the scalar networks, in the MD we see a
consistent nearest-neighbor anti-correlation for all pressures.}
\label{fig:md_pcF}
\end{figure}

Here we discuss the results of our MD simulations.
Fig.\ref{fig:md_pcF} shows $P(F)$, the probability distribution of
vertical forces $F = \vec{{\mathcal{F}}} \cdot \hat{y}$, for MD
systems under various applied pressures $\Pi$.  As with the
scalar model, $F$ has been normalized so that the average vertical
force $\bar{F} = 1$ for each system configuration.  The progression of
$P(F)$ as pressure is increased is very similar both qualitatively and
quantitatively to the crossover from granular to elastic behavior in
the scalar model lattice systems.  We calculate the force-force
correlation values $C_{0}(j)$, shown in Fig.~\ref{fig:md_pcF}(b), by
defining discs to be in plane with a tolerance of $\pm 0.10 a_{D}$ and
$j$ in units of average disc diameter $2 a_{D}$.  In contrast with the
scalar model behavior, the MD systems exhibit a significant
nearest-neighbor anti-correlation for all applied external pressures.
These results for $P(F)$ and $C_{0}(j)$ are independent of whether the
samples are compressed in stages or directly at a fixed pressure
$\Pi$.

\begin{figure}
\centering
\epsfig{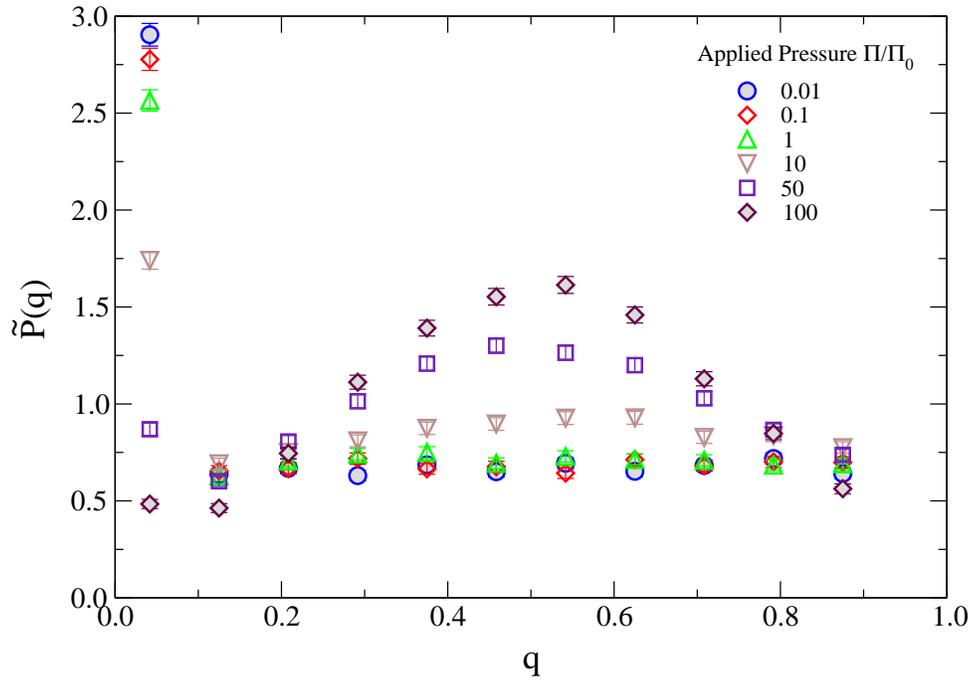}
\caption{Progression of the vertical force fraction $q$ distribution 
$\tilde{P}(q)$ between elastic and granular regimes in MD-generated
packings.  Each $q$ is calculated as the fraction of the vertical force
component that is transferred to the bottom left neighbors of a disc.
At high pressures $\Pi$, we see a peaked form centered at $q=0.5$.  
For decreasing pressure, $\tilde{P}(q)$ flattens out and is roughly
uniform.  The increasing magnitude of the leftward bin with decreasing
pressure $\Pi$ is due to increasing probability of isolated ``rattlers''
and discs with no bottom left neighbors.}
\label{fig:md_pq}
\end{figure}

We define the $q$ value of a disc as the fraction of total vertical
force received from its topward neighbors that is transferred to its
bottom leftward neighbors.  The probability distribution of $q$
values, $\tilde{P}(q)$ is shown in Fig.\ref{fig:md_pq}.  We also
calculate the $q$-$q$ correlation values $\tilde{C}_{0}(j)$ and
$\tilde{C}_{k}(0)$, although the large errors prevent the extraction
of quantitative trends.  Narrowing the statistical errors would be
computationally prohibitive.

\begin{figure}
\centering
\epsfig{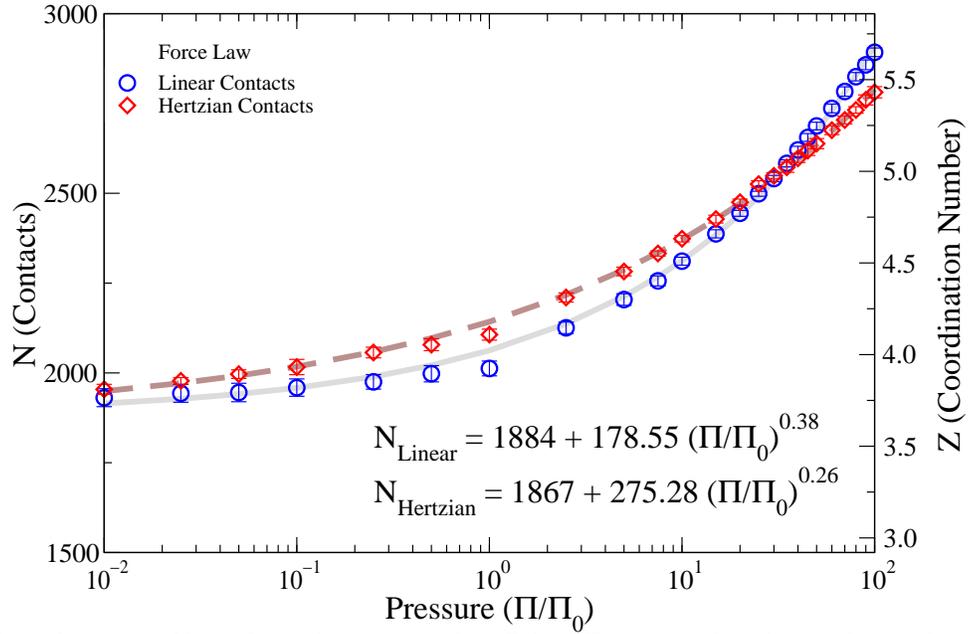}
\caption{Number of contacts $N_{C}$ and coordination number $Z$ for
MD-generated packings of 1024 discs versus externally applied pressure
$\Pi$.  The number of contacts fits roughly to a form $N_{C} = N_{0} + a
(\Pi/\Pi_{0})^{b}$, where $N_{0}$ is the number of contacts in the
zero-force limit.  At low pressures, $N_{0}$ is slightly less than
two contacts per particle because of the presence of ``rattlers'' in
the zero-gravity system.}
\label{fig:md_contacts}
\end{figure}

The number of contacts increases significantly with the pressure, as
shown in Fig.~\ref{fig:md_contacts}.  As the magnitude of the typical
overlap increases, additional contacts are formed.  The number of
contacts at low pressures is below the theoretically predicted average
of $Z = 2 d$\cite{tkachenko99}, where $d$ is the dimension of the
system, because the polydispersity in radii and the lack of gravity
allow for the existence of ``rattlers'' which do not support any of
the external load.

Our results for the Hertzian contact systems are indistinguishable
from those of the linear springs throughout most of the range of
pressures explored.  At higher pressures $\Pi$ ($\Pi \geq 35
\Pi_{0}$), the added stiffness of the Hertzian contacts leads to the
slower narrowing of $P(F)$.

\section{Comparison of results of MD simulations and 
	 of scalar elastic networks}
\label{sec:Comparison}

\begin{figure}
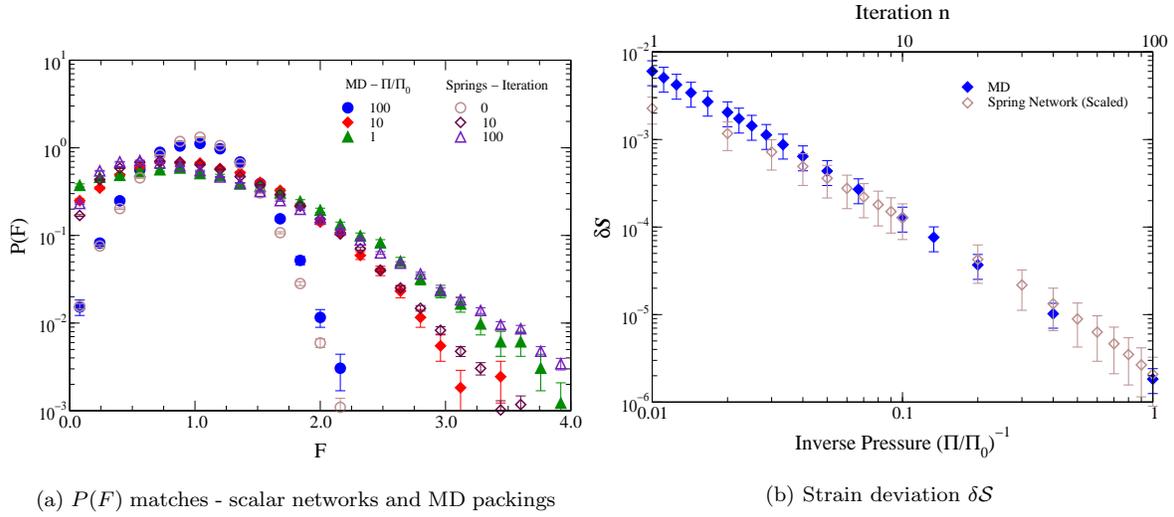

\centering
\begin{tabular}{cc}
\subfigure[$P(F)$ matches - scalar networks and MD packings]
{\epsfig{file=hist_match.eps, width=3in}} &
\subfigure[Strain deviation $\delta {\mathcal{S}}$]
{\epsfig{file=strain_scale.eps, width=3in}}
\end{tabular}
\caption{(a) Matching vertical force probability distributions $P(F)$
for iterated scalar spring networks and MD packings of particles with
a linear force law.  We observe good agreement in the form of $P(F)$
between iteration $n=0$ and $\Pi / \Pi_{0} = 100$, $n=10$ and $\Pi /
\Pi_{0} = 10$, $n=100$ and $\Pi / \Pi_{0} = 1$.  (b) Normalized strain
deviation $\delta {\mathcal{S}}_{N}$ and $\delta {\mathcal{S}}_{MD}$
as a function of actual or equivalent applied pressure.  The matching
of $P(F)$ is used to generate a mapping between iteration values of
the scalar networks and the pressures imposed on the MD systems.  A
simple scaling approach yields $\Pi_{\mathrm{eq}} / \Pi_{0} = 100/n$.
Additionally, $\delta {\mathcal{S}}_{N}$ for the scalar networks is scaled
by a factor of 0.030 to achieve equivalence at $n=10$ ($\Pi / \Pi_{0}
= 10$).}
\label{fig:hist_match}
\end{figure}

Here we compare the behavior observed in the MD simulations and in the
scalar elastic networks.  Because different schemes are used to induce
the granular-elastic crossover in the two systems (iterations in the
scalar networks and external pressure for MD), we need to establish a
common measure to quantify a system's position within the crossover
region.  As the evolution of the probability distribution of vertical
forces, $P(F)$, is qualitatively and quantitatively similar in the
network model and in the MD simulations, we use matches in its form to
establish a relationship between iteration number $n$ and applied
pressure $\Pi$.  Fig.~\ref{fig:hist_match}(a) shows matches in form
between linear-force-law MD packings and iterated scalar network
systems for $\Pi / \Pi_{0} = 100$ and iteration $n = 0$, $\Pi /
\Pi_{0} = 10$ and $n = 10$, and $\Pi / \Pi_{0} = 1$ and $n = 100$
systems.  From these matchings, we map the iteration number $n$ in the
scalar networks to the equivalent applied pressure
$\Pi_{\mathrm{eq}}(n)$ in the MD using the simple
scaling~\cite{footnote3}:
\begin{equation}
\label{eq:scale}
\frac{\Pi_{\mathrm{eq}}(n)}{\Pi_{0}} = \frac{100}{n}.
\end{equation}

We perform a check on this proposed scaling by considering the
analogous quantities of deviation from constant strain in scalar
systems $\delta{\mathcal{S}}_{N}$, given by Eq.~\ref{eq:deviation},
and deviation from the infinitely hard, zero-deformation limit in MD
systems calculated by
\begin{equation}
\delta{\mathcal{S}}_{MD} \equiv
\frac{1}{N_{C}} \sum_{(i,j)} 
  \frac{|\delta r_{i,j}|^{2}}{(a_{i} + a_{j})^{2}}
\end{equation}
where $N_{C}$ is the total number of contacts and the sum is over
pairs of discs $i$ and $j$ in contact.  We match $\delta
{\mathcal{S}}$ values for $\Pi / \Pi_{0} = 10$ and $n = 10$ by scaling
the square deviation for the scalar network systems by a constant
factor of 0.030.  Fig.~\ref{fig:hist_match}(b) shows that this scaling
yields reasonable agreement between $\delta {\mathcal{S}}_{N}$ and
$\delta {\mathcal{S}}_{MD}$ over the crossover region.

\begin{figure}
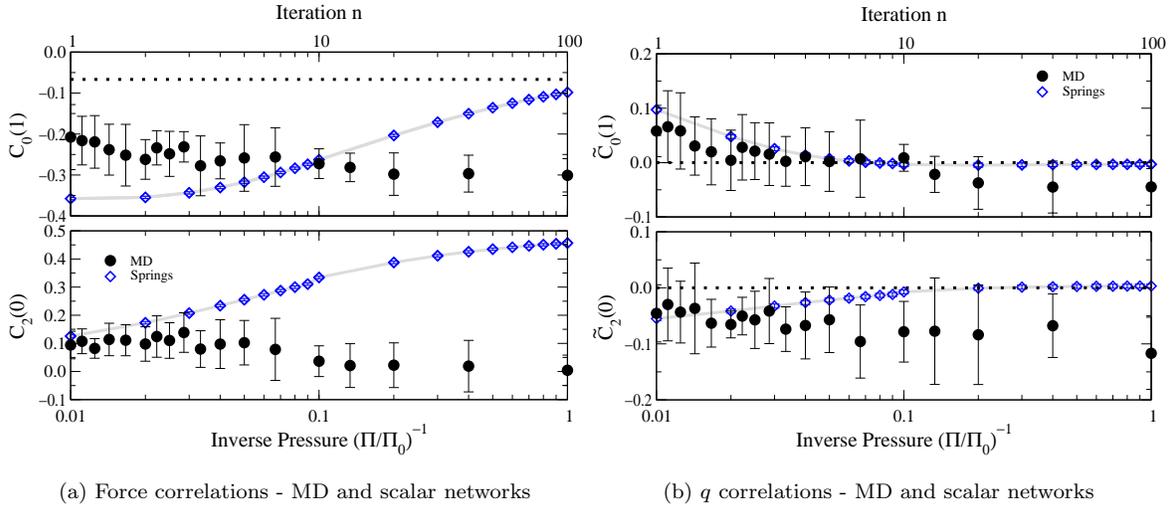

\centering
\begin{tabular}{cc}
\subfigure[Force correlations - MD and scalar networks]
{\epsfig{file=comp_cF.eps, width=3in}} &
\subfigure[$q$ correlations - MD and scalar networks]
{\epsfig{file=comp_cq.eps, width=3in}}
\end{tabular}
\caption{(a) Comparison of nearest-neighbor in-plane and vertical
force-force correlation function values $C_{0}(1)$ and $C_{2}(0)$
between MD and scalar spring networks at various pressures $\Pi$ and
iterations $n$.  We see qualitative and quantitative differences in
behavior.  MD systems exhibit a significant in-plane nearest-neighbor
anti-correlation throughout the crossover region.  This contrasts with
the decrease seen in the scalar networks as it approaches the granular
limit (large iteration $n$).  Strong vertical correlations develop in
the scalar networks but not in the MD systems.  (b) Nearest-neighbor
in-plane and vertical $q$-$q$ correlation function values
$\tilde{C}_{0}(1)$ and $\tilde{C}_{2}(0)$.  The sizeable errors in the
MD values allows for only qualitative comparisons.  Both systems
exhibit similar trends for in-plane nearest-neighbor correlation
values.  However, qualitative differences exist for the behavior of
vertical correlations: the MD systems display a constant
anti-correlation throughout the crossover region while the scalar
networks exhibit an anti-correlation in the elastic regime
which decays rapidly to uncorrelated behavior as the granular limit is
approached.}
\label{fig:correlation_comp}
\end{figure}

In contrast to the agreement in the trends of $P(F)$, qualitative
differences exist between the scalar network model and the MD
simulations in spatial correlation function values $C_{j}(k)$.
Fig.~\ref{fig:correlation_comp}(a) shows the nearest-neighbor in-plane
and vertical force-force correlation values, $C_{0}(1)$ and
$C_{2}(0)$, for the crossover between elastic and granular regimes.
While the MD systems exhibit a significant in-plane nearest-neighbor
anti-correlation throughout the crossover, a decrease in its magnitude
is seen in the scalar networks as the systems change from elastic to
granular.  MD systems do not exhibit strong vertical correlations, in
contrast with the scalar networks whose $C_{2}(0)$ value increases
significantly as the granular limit is approached.

The large statistical uncertainties in our $q$-$q$ correlation
functions for MD systems restrict us to making only qualitative
behavior descriptions.  The trend for in-plane nearest-neighbor
correlation behavior $\tilde{C}_{0}(1)$ in both systems is similar.
However, qualitative differences exist for vertical correlation value
$\tilde{C}_{2}(0)$: the MD systems display consistent anti-correlation
behavior, while the scalar networks display anti-correlation behavior
in the elastic regime which decays rapidly to uncorrelated behavior as
the granular limit is approached.

Our work indicates that experiments on granular media at high pressures
should yield a force histogram that differs qualitatively from that
observed at lower pressures.  Experiments by Howell
\textit{et al.}~\cite{howell99} as well as experiments and simulations
by Makse \textit{et al.}~\cite{makse00} are in qualitative agreement
with this result.  Howell \textit{et al.}~\cite{howell99} control the
transition between granular and elastic behavior of slowly sheared
systems in a 2-D Couette geometry by varying the packing fraction
$\gamma$ within a range $0.77 \leq \gamma \leq 0.81$.  The average
force/length on a particle increases with $\gamma$.  For lower values
of $\gamma$, the distribution of large stresses is asymptotically
exponential, while the distribution of stresses has a gaussian form at
higher packing fractions $\gamma$.  Makse \textit{et
al.}~\cite{makse00} apply increasing pressure to three-dimensional
packings of spherical glass beads to achieve the crossover between
granular and elastic behavior and also perform MD simulations on 3-D
systems.  Makse \textit{et al.}~observe a crossover in the force
histogram $P(F)$ in a pressure range that is consistent with our 2-D
MD results.

An interesting question is whether the persistent in-plane
nearest-neighbor anti-correlation in the forces that is observed in
the MD simulations is present in experimental systems.  Mueth
$\textit{et al.}$~\cite{mueth98} do not find evidence of correlations
between different sites in the same horizontal layer; any
nearest-neighbor anti-correlation in the experiment is smaller than
the experimental resolution.  However, they measure a different
correlation function, $K_{1}(r)$, defined as
\begin{equation}
K_{1}(r) = 
\frac{\displaystyle \sum_{i=1}^{N_{B}} \sum_{j=i+1}^{N_{B}} 
      \delta(r_{ij}-r) f_{i} f_{j}}
     {\displaystyle \sum_{i=1}^{N_{B}} \sum_{j=i+1}^{N_{B}} \delta(r_{ij}-r)},
\end{equation}
where the sums are over the $N_{B}$ particles in the bottom layer,
$f_{i}$ is the force at position $r_{i}$ in the bottom layer and
$r_{ij} = |\vec{r}_{i} - \vec{r}_{j}|$.  Calculation of $K_{1}(r)$
from the numerical data for our MD simulations yields values of the
correlation function that are smaller than the error bars in the
experiment.  Comparison with Ref.~\cite{mueth98} is necessarily
qualitative since the experiments measure the properties at the
surface of a 3-D packing while our MD results are calculated using
numerical data from the bulk of a 2-D system.

\section{Discussion}
\label{sec:Discussion}
We have investigated the crossover between elastic and granular stress
transmission in both a 2-D scalar lattice model and in molecular
dynamics simulations of slightly polydisperse discs.  The evolution of
$P(F)$, the probability distribution of stresses, is very similar in
the lattice model and in the MD.  However, the behavior of the spatial
correlation functions for stress, $C_{k}(j)$, differs qualitatively.

Our investigations of the scalar model have several implications for
the development of granular media models.  First, we have shown that
implementing a local constraint can convert an elastic network to a
$q$-model.  This constraint has the natural physical interpretation
that the strain in the system must be uniform; it is plausible that
rearrangements would prevent strain gradients from forming.  Second,
implementing this constraint to increasing accuracy causes the force
histogram $P(F)$ to evolve in a manner similar to that observed in the
MD simulation as the pressure is decreased.  $P(F)$ has a tail
consistent with exponential decay at large forces in the granular
limit, while the $P(F)$ for the highly compressed system is much
narrower and decays more quickly at large forces.  We note that
implementing a non-tensile force constraint alone, as in
Ref.~\cite{sexton99}, yields gaussian decay in $P(F)$ at large forces
even at the lowest pressures, in qualitative disagreement with the MD
results of us and others~\cite{radjai96,radjai97,makse00}.

While this success in describing the evolution of the force histogram
and the scalar model's simplicity in both formulation and
implementation make it an attractive platform for the study of media
models, the discrepancy in the behavior of the correlation function
behavior with the MD simulation results needs to be addressed.  The
scalar model assumes explicitly that in the granular regime the stress
redistribution fractions $q$ at different sites are uncorrelated.  The
extent to which this condition is valid needs to be examined in more
detail.  Spatial correlations of the $q$'s can strongly affect the
probability distribution of stress $P(F)$~\cite{claudin97, nicodemi98}
but the degree to which these correlations exist in real packings has
not been settled.  A possible source of spatial correlations in the
$q$'s is the constraint that non-tensile vector forces must be
balanced.  However, vector generalization of $q$-model systems that
have been proposed to date have required arbitrary constraints to be
imposed to limit the scale of stress components perpendicular to the
direction of applied force~\cite{footnote2}.  Clarification of the
roles of vector force balance and contact formation is key to
identifying and characterizing the processes governing stress
transmission beyond those that have been implemented in the scalar
model.

In conclusion, we have shown that similarities exist in the evolution
of the probability distribution of stresses $P(F)$ in the crossover
between elastic and granular regimes for a scalar lattice model and MD
simulations of slightly polydisperse discs.  However, the systems
exhibit qualitative differences in the two-point force correlation
function $C_{k}(j)$.  Further investigation of the systematic
influences leading to the spatial correlations between forces is
necessary for the development of a successful model of stress
transmission in granular media.

\section*{Acknowledgements}
\label{sec:Acknowledge}

We thank Alexei Tkachenko and Tom Witten for a key suggestion and Rick
Clelland, Heinrich Jaeger, Dan Mueth, Sid Nagel, Joshua Socolar, and
Bob Behringer for useful conversations.  This research was supported by
the MRSEC program of the NSF and by the Petroleum Research Fund of the
American Chemical Society.

\appendix

\section{Finite-size correction to correlation function calculation}
\label{app:Correlation}
In the $q$-model in the limit of infinite size, forces at
different sites in the same layer are completely uncorrelated.
In a system of finite transverse extent, the requirements that
the total force through every layer is identical
and there are no tensile forces lead to a finite-size
correction to the correlation function.  This appendix discusses
this correction.

We characterize the correlations between force fluctuations
on sites in the same row using the correlation function
\begin{equation}
    C_{0}(j) = 
	  \frac{1}{L_{y}} \sum_{l=1}^{L_{y}} \left(
	  \frac{\displaystyle 
                \sum_{m=1}^{L_{x}} \delta F_{l,m} \delta F_{l,m+j}}
	       {\displaystyle \sum_{m=1}^{L_{x}} \delta F_{l,m}^2}\right),
\end{equation}
where $\delta F_{l,m} = F_{l,m} - \bar{F}$ is the deviation of the force
at a site in row $l$ and column $m$ from the average force $\bar{F}$.  For
the $q$-model with a uniform distribution of $q$'s, in a system of
infinite transverse extent this correlation function
is~\cite{coppersmith96}
\begin{equation}
    C_{0}(j) = \left\{ \begin{array}{ll}
                      1 & j = 0 \\
                      0 & j \neq 0 
                      \end{array}
               \right. .
\end{equation}
This result follows from the fact that $P(F_{\alpha}, F_{\beta})$, the
probability that the force through node $\alpha$ is $F_{\alpha}$ and
the force through node $\beta$ at the same horizontal level is
$F_{\beta}$, is factorizable:
\[  P(F_{\alpha}, F_{\beta}) = P(F_{\alpha}) P(F_{\beta}) ,\]

As a result, for $\alpha \neq \beta$
\begin{eqnarray}  
    \langle \delta F_{\alpha} \delta F_{\beta} \rangle & = &
	  \lim_{L_{x}, L_{y} \rightarrow \infty} 
          \frac{1}{L_{y} L_{x}} 
	  \sum_{l=1}^{L_{y}}
          \sum_{m=1}^{L_{x}} \delta F_{l,m} \delta F_{l,m+j} \\
    & = & \int_{0}^{\infty} \int_{0}^{\infty} 
      dF_{\alpha} dF_{\beta} \ \delta F_{\alpha} \ \delta F_{\beta}
                             \ P(F_{\alpha}, F_{\beta}) \nonumber \\
    & = & \int_{0}^{\infty} \int_{0}^{\infty} 
          dF_{\alpha} dF_{\beta} \ \delta F_{\alpha} \ \delta F_{\beta}
                                 \ P(F_{\alpha}) P(F_{\beta}) \nonumber\\
    & = & \left(\int_{0}^{\infty} 
                dF_{\alpha} \ \delta F_{\alpha} \ P(F_{\alpha}) 
          \right)
          \left(\int_{0}^{\infty} 
                dF_{\beta} \ \delta F_{\beta} \ P(F_{\beta})
          \right) \nonumber\\
    & = & \left[ \int_{0}^{\infty} dF \delta F \ P(F) \right]^{2} \nonumber\\
    & = & \left[ \int_{0}^{\infty} dF (F - \bar{F}) \ P(F) \right]^{2} 
          \nonumber\\
    & = & \left[ \bar{F} - \bar{F} \right]^{2} \nonumber\\
    & = & 0 \nonumber.
\end{eqnarray}

On a lattice of finite width ($L_{x}$ sites),
the multipoint force probability distribution function
must be consistent with the facts that first, the total force
down every layer is fixed, and second, no force is
negative.  This implies that
\begin{list}{$\bullet$}{}
  \item The maximum force on any node in any layer cannot be
        larger than $F_{\mathrm{max}} = L_{x} \bar{F}$, and

  \item The force $F_{\alpha}$ at a node $\alpha$ contributes
        to the total force along a layer and hence affects the 
        sum of the forces through the remaining sites in the layer.
        Defining $\tilde{F}$ as the average force through all the
        sites in the layer other than site $\alpha$, we have
        \begin{eqnarray} 
          \tilde{F} & = & \frac{L_{x} \bar{F} - F_{\alpha}}{L_{x} - 1} \\
                    & = & \bar{F} - \frac{F_{\alpha} - \bar{F}}{L_{x} - 1} 
                          \nonumber \\
                    & = & \bar{F} - \frac{\delta F_{\alpha}}{L_{x} - 1} .
                          \nonumber 
        \end{eqnarray}
\end{list}

Assuming that the only correlations present in the finite
system are those required to satisfy these conditions,
the joint probability distribution in the system with
finite $L_x$ can again be written
\begin{equation}
      P(F_{\alpha}, F_{\beta}) = \frac{1}{\bar{F} \tilde{F}}
      P(\frac{F_{\alpha}}{\bar{F}}) \tilde{P}(\frac{F_{\beta}}{\tilde{F}}) ,
\end{equation}
but now the distributions
are subject to the constraints
\begin{mathletters}
\begin{eqnarray}
  \int_{0}^{L_{x}} P(\omega) d\omega = 1 &, \ \ \ &
  \int_{0}^{L_{x}} \omega P(\omega) d\omega = 1 \\
  \int_{0}^{L_{x} - 1} \tilde{P}(\nu) d\nu = 1 &, \ \ \ &
  \int_{0}^{L_{x} - 1} \nu \tilde{P}(\nu) d\nu = 1 ~.
\end{eqnarray}
\end{mathletters}
In the limit of $L_{x} \rightarrow \infty$, we expect corrections to
$P(F)$ to be of order $1/L_x$ as the fluctuations in the forces at the
sites are of order unity.  As we will see, with the assumptions that
we have made, the finite size correction to the correlation function
does not depend on the form of the probability distribution $P$.
Taking the new constraints into account, for $\alpha \neq \beta$ we
have
\begin{eqnarray}
    \left< \delta F_{\alpha} \delta F_{\beta} \right> & = & 
      \int_{0}^{L_{x}} dF_{\alpha} 
      \int_{0}^{L_{x} - F_{\alpha}} dF_{\beta} 
        \ \delta F_{\alpha} \ \delta F_{\beta}
        \ P(F_{\alpha}, F_{\beta}) \\
    & = & \int_{0}^{L_{x}} dF_{\alpha} 
          \int_{0}^{L_{x} - F_{\alpha}} dF_{\beta} 
          \ (F_{\alpha} - \bar{F}) (F_{\beta} - \bar{F})
          \left( \frac{1}{\bar{F} \tilde{F}} 
                 P\left(\frac{F_{\alpha}}{\bar{F}}\right) 
                 \tilde{P}\left(\frac{F_{\beta}}{\tilde{F}}\right) 
          \right) \nonumber\\
    & = & \bar{F}^{2} 
          \int_{0}^{L_{x}} d\left(\frac{F_{\alpha}}{\bar{F}}\right)
          \left(\frac{F_{\alpha}}{\bar{F}} - 1\right)
          P\left(\frac{F_{\alpha}}{\bar{F}}\right)
          \frac{\tilde{F}}{\bar{F}}
          \int_{0}^{L_{x} - F_{\alpha}} 
          d\left(\frac{F_{\beta}}{\tilde{F}}\right)
          \left(\frac{F_{\beta}}{\tilde{F}} - \frac{\bar{F}}{\tilde{F}}\right)
          \tilde{P}\left(\frac{F_{\beta}}{\tilde{F}}\right) \nonumber\\
    & = & \bar{F}^{2} 
          \int_{0}^{L_{x}} d\left(\frac{F_{\alpha}}{\bar{F}}\right)
          \left(\frac{F_{\alpha}}{\bar{F}} - 1\right)
          P\left(\frac{F_{\alpha}}{\bar{F}}\right)
          \frac{\tilde{F}}{\bar{F}}
          \int_{0}^{L_{x} - 1}
          d\nu \left(\nu - \frac{\bar{F}}{\tilde{F}}\right) \tilde{P}(\nu) 
          \nonumber\\
    & = & \bar{F}^{2} 
          \int_{0}^{L_{x}} d\left(\frac{F_{\alpha}}{\bar{F}}\right)
          \left(\frac{F_{\alpha}}{\bar{F}} - 1\right)
          P\left(\frac{F_{\alpha}}{\bar{F}}\right)
          \left[
            \frac{\tilde{F}}{\bar{F}} 
            \int_{0}^{L_{x} - 1} d\nu \nu \tilde{P}(\nu) -
            \int_{0}^{L_{x} - 1} d\nu \tilde{P}(\nu)
          \right] \nonumber\\
    & = & \bar{F}^{2}
          \int_{0}^{L_{x}} d\left(\frac{F_{\alpha}}{\bar{F}}\right)
          \left(\frac{F_{\alpha}}{\bar{F}} - 1\right)
          P\left(\frac{F_{\alpha}}{\bar{F}}\right) \times \nonumber\\
    &   & \ \ \ \ 
          \left[
            \left(1 - 
                  \frac{1}{L_{x} - 1} \left(\frac{F_{\alpha}}{\bar{F}}\right)
            \right)
            \int_{0}^{L_{x} - 1} d\nu \nu \tilde{P}(\nu) -
            \int_{0}^{L_{x} - 1} d\nu \tilde{P}(\nu)
          \right] \nonumber\\
    & = & \bar{F}^{2}
          \int_{0}^{L_{x}} d\omega (\omega - 1) P(\omega) 
          \left[\left(1 - \frac{\omega - 1}{L_{x} - 1}\right)
            \int_{0}^{L_{x} - 1} d\nu \nu \tilde{P}(\nu) -
            \int_{0}^{L_{x} - 1} d\nu \tilde{P}(\nu)
          \right] \nonumber\\
    & = & \bar{F}^{2}
          \int_{0}^{L_{x}} d\omega (\omega - 1) P(\omega) 
          \left[\int_{0}^{L_{x} - 1} d\nu (\nu - 1) \tilde{P}(\nu) -
                \frac{\omega - 1}{L_{x} - 1} 
                \int_{0}^{L_{x} - 1} d\nu \nu \tilde{P}(\nu)
          \right] \nonumber\\
    & = & \bar{F}^{2} 
          \left[\int_{0}^{L_{x}} d\omega (\omega - 1) P(\omega)
                \int_{0}^{L_{x} - 1} d\nu (\nu - 1) \tilde{P}(\nu) - 
          \right. \nonumber\\
    &   & \ \ \ \ \ \ \ \ \ \ \ 
          \left.
                \frac{1}{L_{x} - 1} 
                \int_{0}^{L_{x}} d\omega (\omega - 1)^{2} P(\omega)
                \int_{0}^{L_{x} - 1} d\nu \nu \tilde{P}(\nu)
          \right] \nonumber\\
    & = & -\frac{\bar{F}^{2}}{L_{x} - 1} 
                \int_{0}^{L_{x}} d\omega (\omega - 1)^{2} P(\omega) .
          \nonumber
\end{eqnarray}
As the correlation function is normalized with respect to the 
average fluctuation size,
\begin{eqnarray}
  C_{0}(j \neq 0) & = & \lim_{L_{y} \rightarrow \infty}
	                \frac{1}{L_{y}} \sum_{l=1}^{L_{y}} \left(
                        \frac{\displaystyle 
                \sum_{m=1}^{L_{x}} \delta F_{l,m} \delta F_{l,m+j}}
	       {\displaystyle \sum_{m=1}^{L_{x}} \delta F_{l,m}^2}\right) \\
	          & = & \frac{\left< \delta F_{\alpha} \delta
	                F_{\beta} \right>_{\alpha \neq \beta}}
   	                  {\left< \delta F_{\alpha}^{2} \right>} \nonumber \\
                  & = & \frac{-\frac{\bar{F}^{2}}{L_{x} - 1}
                               \int_{0}^{L_{x}} d\omega 
                                (\omega - 1)^{2} P(\omega)}
                             {\bar{F}^{2} \int_{0}^{L_{x}} d\omega 
                               (\omega - 1)^{2} P(\omega)} \nonumber\\
                  & = & -(L_{x} - 1)^{-1} \nonumber
\end{eqnarray}
which is independent of the form of the $P(F)$.
As $L_x \rightarrow \infty$, $C_{0}(j \neq 0) \rightarrow 0$,
as expected.

\bibliographystyle{prsty}

\newpage


\clearpage

\clearpage

\clearpage

\clearpage

\clearpage

\clearpage

\clearpage

\clearpage

\clearpage

\clearpage

\clearpage

\clearpage

\clearpage

\clearpage

\end{document}